\documentclass[10pt,journal,twoside]{IEEEtran}
\usepackage{mathtools}
\usepackage{amsthm}

\usepackage{float}
\usepackage[colorlinks,linkcolor=red,anchorcolor=red,citecolor=red]{hyperref}
\usepackage[pdftex]{graphicx}
\graphicspath{{.}}

\usepackage{subfig}
\usepackage{subfloat}
\usepackage{multicol}
\usepackage{paralist}
\usepackage{bm}
\usepackage{amsfonts}
\usepackage{url}
\usepackage{multirow}
\usepackage{cases}

\usepackage{booktabs}
\usepackage{threeparttable}
\usepackage{color}
\usepackage{xcolor}
\usepackage{colortbl,booktabs}

\makeatletter
\newif\if@restonecol
\makeatother

\usepackage[linesnumbered,ruled,vlined]{algorithm2e}
\usepackage{algpseudocode}
\usepackage{amsmath}

\begin{document}

\title{A Blockchain-enabled Trustless Crowd-Intelligence Ecosystem on Mobile Edge Computing}

\author{Jinliang~Xu, Shangguang~Wang,~\IEEEmembership{Senior Member,~IEEE,} ~Bharat K. Bhargava,~\IEEEmembership{IEEE Fellow},
        ~Fangchun~Yang,~\IEEEmembership{Senior Member,~IEEE}
\IEEEcompsocitemizethanks{\IEEEcompsocthanksitem Jinliang Xu, Shangguang Wang, and Fangchun Yang are with the State Key Laboratory of Networking and Switching Technology, Beijing University of Posts and Telecommunications.
\protect E-mail: {jlxu@bupt.edu.cn; sgwang@bupt.edu.cn; fcyang@bupt.edu.cn},
}
\IEEEcompsocitemizethanks{\IEEEcompsocthanksitem Bharat K. Bhargava is with the Department of Computer Science, Purdue
University, West Lafayette, IN 47906, USA.
\protect E-mail: {bbshail@purdue.edu},
}
\thanks{\bf This article has been accepted for publication in a future issue of this journal, but has not been fully edited. Content may change prior to final publication. Citation information: DOI 10.1109/TII.2019.2896965, IEEE
Transactions on Industrial Informatics
}
}
\markboth{IEEE Transactions on Industrial Informatics,  VOL. x, NO. x, xx/xx 2019}{XU et al.: A Blockchain-enabled Trustless Crowd-Intelligence Ecosystem on Mobile Edge Computing}%

\maketitle

\begin{abstract}
Crowd-intelligence tries to gather, process, infer and ascertain massive useful information by utilizing the intelligence of crowds or distributed computers, which has great potential in  Industrial Internet of Things (IIoT). A crowd-intelligence ecosystem involves three stakeholders, namely the platform, workers (e.g., individuals, sensors or processors), and task publisher. The stakeholders have no mutual trust but interest conflict, which means bad cooperation of them. Due to lack of trust, transferring raw data (e.g., pictures or video clips) between publisher and workers requires the remote platform center to serve as a relay node, which implies network congestion. First we use a reward-penalty model to align the incentives of stakeholders. Then the predefined rules are implemented using blockchain smart contract on many edge servers of the mobile edge computing network, which together function as a trustless hybrid human-machine crowd-intelligence platform. As edge servers are near to workers and publisher, network congestion can be effectively improved. Further, we proved the existence of the only one strong Nash equilibrium, which can maximize the interests of involved edge servers and make the ecosystem bigger. Theoretical analysis and experiments validate the proposed method respectively.

\end{abstract}
\IEEEpeerreviewmaketitle

\section{Introduction}\label{sec:introduction}
\IEEEPARstart{C}{}rowd-intelligence is to gather, process, infer and ascertain massive useful information by utilizing the intelligence of crowds or computers, whereby a publisher broadcasts massive tasks to lots of semi-skilled workers to obtain reliable answers \cite{ooi2014contextual}. It is widely used in  knowledge collecting (e.g. mobile crowdsensing \cite{guo2017emergence}), decision making (e.g., tagging of machine learning training dataset \cite{radu2013error}), etc. {Industrial Internet of Things (IIoT) has the properties of high distribution, high-frequency activity, and is near to massive mobile users, which makes itself a great potential for crowd-intelligence \cite{zhang2017crowdsourcing,fernandes2015iot}}.  The number of tasks to be completed is too large to complete on time for limited number of professionals, or the tasks are too difficult to be auto-processed well just by computer programs \cite{radu2013error,gao2015on,ren2015exploiting}. A crowd-intelligence ecosystem involves three stakeholders, namely  platform, workers , and publisher that can publish tasks to workers.

These three stakeholders may have no mutual trust and their interests conflict with each other \cite{gao2015on,xu2018reward,jiang2017trust}. For example, workers try to get more payment from the publisher with low quality works. It is not easy to align incentives of stakeholders so that they may collaborate smoothly. This harms the long term development of the whole crowd-intelligence ecosystem, e.g., answers of low quality, unnecessary high cost paid by publisher to workers, delayed completion of the tasks (i.e., exceeding the time constraint set by the publisher), and weakening platform (i.e., loss of professional workers, and poorly prespared workers flooding into it). Due to the lack of mutual trust among stakeholders, the raw data for a task (e.g., pictures or video clips that are published by the publisher or submitted by workers \cite{guo2017emergence,fan2017crowdsourced,radu2013error}) cannot be transferred directly between publisher and workers. The remote cloud center (RC) of the centralized crowd-intelligence platform must serve as the third party guarantee and the intermediary node of the data transferring network path. As raw data transferring plays a great role in a normal crowd-intelligence ecosystem, trust problem may result in excess bandwidth and delayed response \cite{wollschlaeger2017future,lindgren2017end,khan2014survey}. This needs to be avoided in time critical use cases, e.g., automobile navigation \cite{fan2017crowdsourced, kang2018blockchain, liu2018blockchain} or disaster recovery \cite{satyanarayanan2013role}. Finally, a crowd-intelligence ecosystem is short of the available workers to complete growing number and variety of tasks and this leads to low quality of final answers and delayed completion of the complete tasks.

\begin{figure}[!t]
\centering
\setlength{\belowcaptionskip}{-0.5cm}
\begin{center}
\includegraphics*[width=0.75\linewidth]{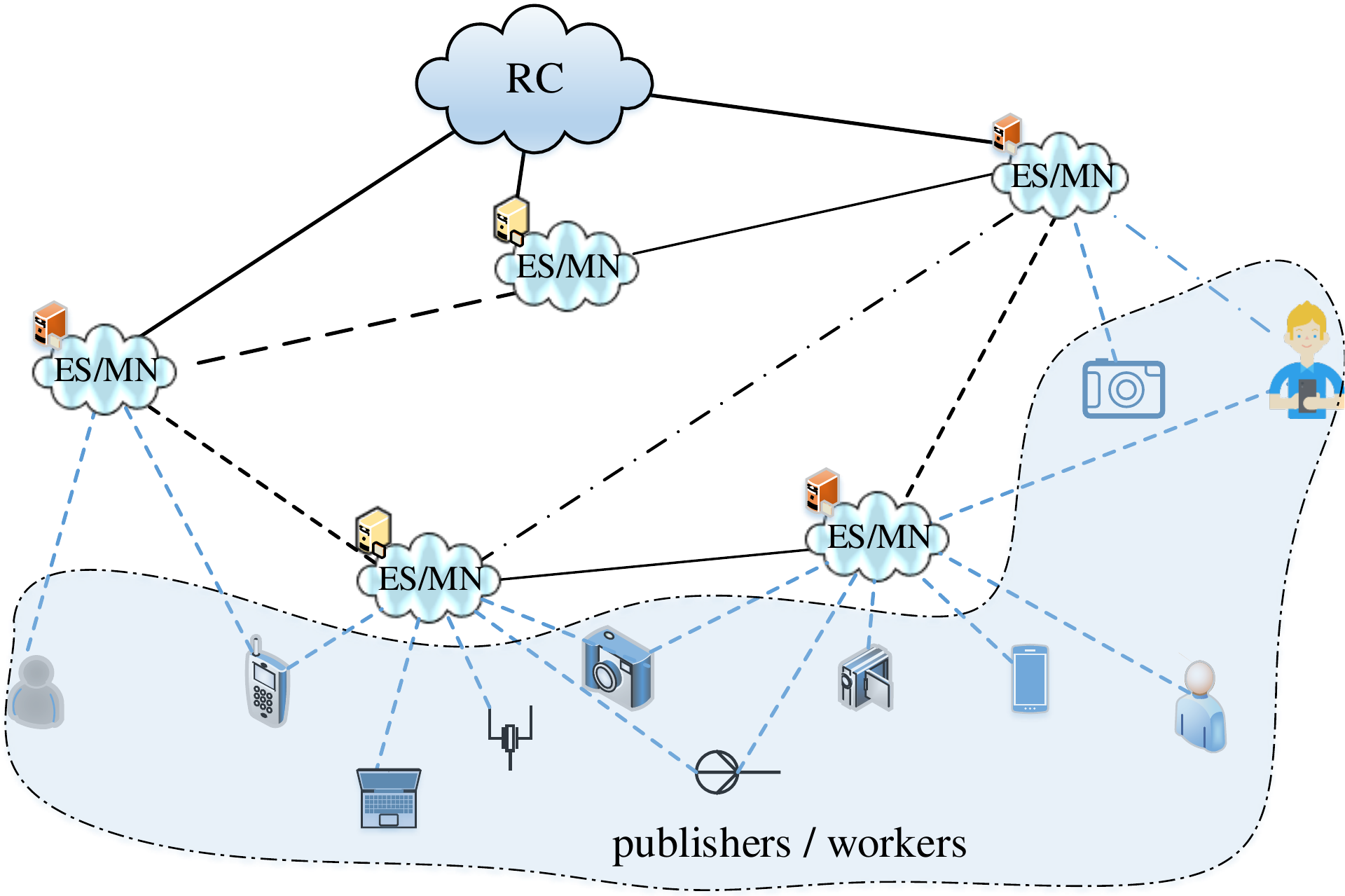}
\caption{The proposed blockchain-enabled trustless crowd-intelligence ecosystem. It is decentralized and implements a blockchain smart contract and operates on edge servers$/$blockchain masternodes (ES$/$MNs) of a mobile edge computing network. Both human beings and machines can serve as workers or publishers.}
\label{fig:CrowdIntelligenceEcosystem}
\end{center}
\end{figure}

We solve the above-mentioned problems from the following three aspects, and build them into an integrated method named blockchain-enabled trustless crowd-intelligence ecosystem operated on mobile edge computing network (Fig. \ref{fig:CrowdIntelligenceEcosystem}, {note that remote cloud (RC) is not obligatory in general. But without RC, the proposed decentralized crowd-intelligence network still works}.):
\begin{inparaenum}[i)]
\item {\bf incentive problem}. An incentive-compatible and efficient incentive model is particularly important for a decentralized system. We build a reward-penalty model to align incentives of the three stakeholders. We reward or punish workers according to the quality of their committed answers instead of only positive payment \cite{fan2017crowdsourced,radanovic2013robust}. We develop a family of appropriate reward-penalty function couples, by which the amounts of reward and penalty can be computed based on the workers' committing beliefs without destroying the incentive compatibility property of the ecosystem. Different from a single reward-penalty function couple or rewarding only, this model provides a more flexible solution to manage the interests of three stakeholders, and help to reduce latency, improve quality, and do good to platform evolution of the crowd-intelligence ecosystem \cite{fan2017crowdsourced, radanovic2013robust}.
\item {\bf trust problem}. A worker's performance history determines the amounts of her reward$/$penalty in completing tasks. And the predefined management rules$/$cooperation standards define how the platform works. It is of important to avoid malicious activity or tampering from the bad minorities, and win the trust of stakeholders. We propose to treat it with blockchain smart contract. The smart contract is copied to multiple edge servers (ES) of mobile edge computing and executed. The proposed crowd-intelligence platform is decentralized using blockchain technology and is not hosted on a centralized RC. The ESs are owned by many individuals and behave as blockchain masternodes (MNs) \cite{kim2018secure}. Without the agreement of more than a half MNs, the smart contract cannot be changed arbitrarily \cite{underwood2016blockchain}. The stakeholders can collaborate without trusting each other. Specifically, the raw data transferring between publisher and workers can proceed with the nearest ES/MNs serving as intermediary nodes, which can avoid excess bandwidth and delayed response.  The ES/MNs charge a small fees from the publisher and workers to keep the security state of the blockchain, and earn money from it \cite{kim2018secure, underwood2016blockchain}.
With the help of blockchain smart contract, the involved ESs can maximize profits with strong Nash equilibrium \cite{shubik1970game} and the proposed ecosystem can attract more ESs  to serve as blockchain masternodes.
\item {\bf worker shortage problem}. When the trust problem is solved, a trustless hybrid human-machine crowd-intelligence platform can be built, where both of human beings and machines (including sensors or processors) are able to serve as workers or publishers without mutual trust. Specifically, human workers perform well in highly complex tasks (i.e., crowdsourcing \cite{chittilappilly2016survey}), while machine workers can do better in realtime high-frequency tasks (i.e., crowdsensing \cite{fan2017crowdsourced}). This platform can also help to solve the worker shortage problem.
\end{inparaenum}

The rest of this paper is organized as follows:
We introduce how to align the interests of three involved stakeholders using the reward and penalty model in Section \ref{sec:reward-penalty}. Section \ref{sec:trust} shows how to build a trustless crowd-intelligence ecosystem using blockchain smart contract on a mobile edge computing network. In Section \ref{sec:hybrid}, we introduce the hybrid human-machine concept to relieve the worker shortage problem. In Section \ref{sec:results}, experiments on two synthetic datasets are conducted to validate the proposed
crowd-intelligence ecosystem, and the results are supplements
to the proof and analysis in the previous sections. Section \ref{sec:relatedwork} presents the existing related works. And Section \ref{sec:conclusion} draws conclusions and discusses our future work.

\section{Reward-Penalty Model}\label{sec:reward-penalty}
{An incentive-compatible and efficient incentive model is particularly important for a decentralized system. Not like the  traditional centralized
crowd-intelligence system, the proposed decentralized version has no an arbitral authority to settle the disputes among different stakeholders. If the system is incentive-compatible, every stakeholder can achieve the best outcome to themselves just by behaving according to predefined rules, which helps to avoid disputes. If the proposed decentralized crowd-intelligence system is efficient, it will have more advantages over the existing centralized competitors. As a result, a good incentive model that can achieving the above targets is needed.}

The proposed reward-penalty model does not estimate a worker's proficiency or  her reliability on each commit. Instead, it assumes that each worker knows her reliability on a specific task, and its target is to incentive workers to commit the truth. If a worker is assigned a task, she is asked to commit both of task's \emph{type} information (i.e. the answer she selected from the candidates) and \emph{belief value} (i.e. her confidence that measures the probability that her selection will be judged right). Only binary-type task is taken into account here, such that the number of the task's candidate answers is uniformly two, like \emph{Yes} or \emph{No} \cite{witkowski2012peer,xu2018reward,radanovic2013robust}.
\emph{Belief value} is reasonable because that a worker always intentionally or unintentionally estimates in her own mind the probability of giving the right answer  when she is encountered with a decision making task due to tasks' inherent difficulty, different individual experience and professional knowledge, etc.  \cite{vullioud2016confidence,radanovic2015incentives}.

Some symbols and denotations are as follows. The number of tasks is denoted by $T$, and there are a total of $N$ workers to complete these tasks. The index of a task is denoted by an integer $1\le t\le T$, and a worker is indexed using an integer  $1\le n\le N$.
The binary-type space of tasks is denoted as $\{-1,+1\}$.  We use $s_t$ to denote the worker set that completed task $t$.
When worker $n$ completes task $t$, we use $a_{n,t}$ to represent her committing type for this task. The \emph{belief} value is in range $[0.5,1]$, which means that the committed type is more suitable than the other one. $x$ denotes the committed belief value $x$, and the real probability that her committed type is judged right is denoted by $c$. When a worker's committing type is finally judged correct, her reward amount is denoted by the reward function of variable $x$, i.e., $reward(x)>0$, and $penalty(x)>0$ is the penalty function if her committing type is judged as wrong.

\begin{figure}[!t]
\centering
\setlength{\belowcaptionskip}{-0.5cm}
\begin{center}
\includegraphics*[width=0.75\linewidth]{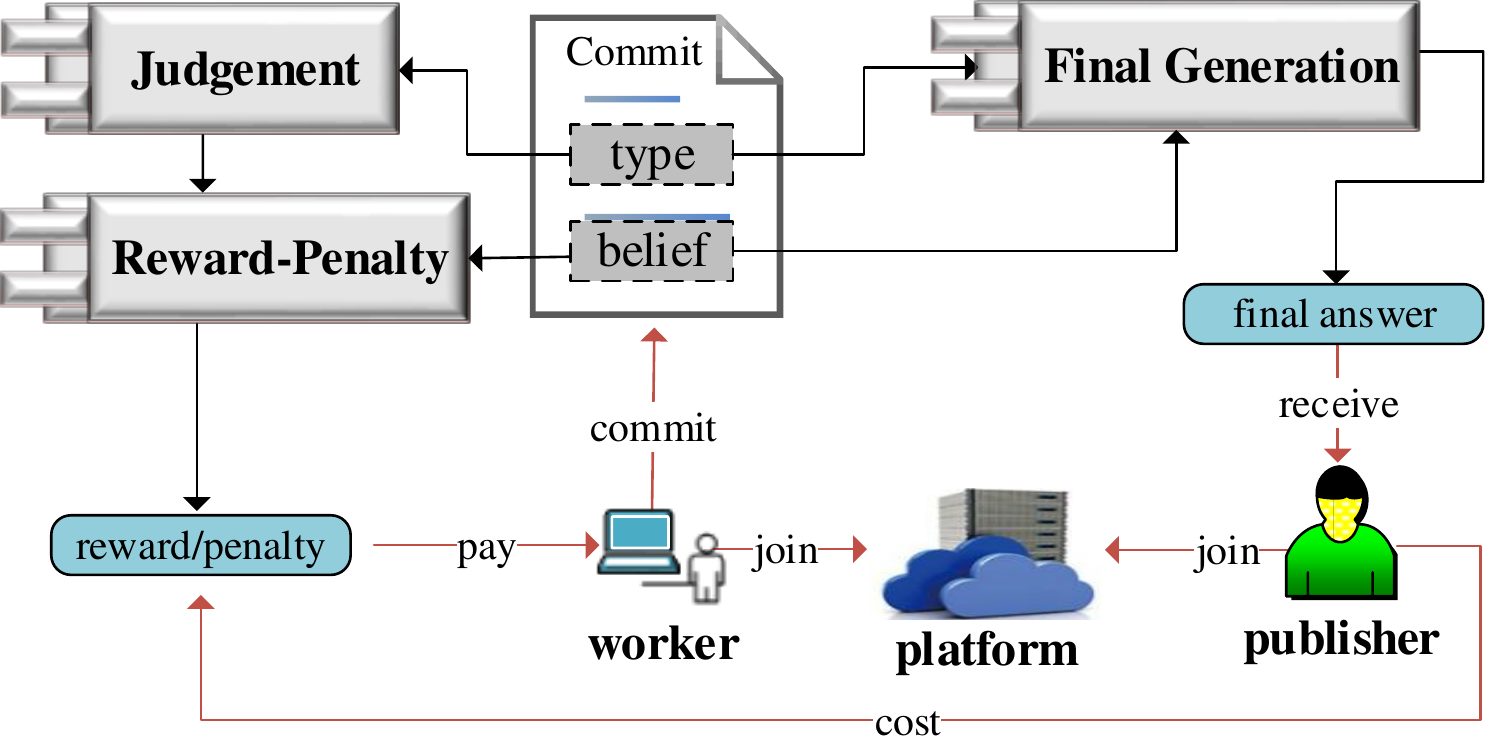}
\caption{Workflow of the reward-penalty model. }
\label{fig:flow}
\end{center}
\end{figure}

The workflow of reward-penalty model is shown in Fig. \ref{fig:flow}, and it contains three modules, namely \emph{Judgement}, \emph{Final Generation}, and \emph{Reward-Penalty}:

\subsection{Judgement Module}
Judgement module uses \emph{benchmark answer} to judge whether a work for a task deserves reward or penalty for a task completion. The \emph{benchmark answer} for task $t$ is defined by the following:
\begin{equation}\label{equ:judgement}
\begin{split}
\widehat{a_t} = &\mathrm{sgn}\left(\sum_{n\in s_t}a_{n,t}\right)
\end{split},
\end{equation}
where $\mathrm{sgn}(x)=-1$ if $x< 0$, and $1$ otherwise.

Note that benchmark answer $\widehat{a_t}$ will not be sent to the publisher as the final answer of a task. It is just used to judge whether or not worker $n$ is due for a reward or penalty for her $a_{n,t}$. Specifically, if $a_{n,t}$ equals to $\widehat{a_t}$, the worker will receive a reward from the publisher, otherwise she will need to pay to the publisher as a penalty.

\subsection{Reward-Penalty Module}
This module computes the how much reward ($reward(x)$) or penalty ($penalty(x)$) should be given. In statistical terms, the type committed by a worker will be judged correct with probability of $c$, and occurrence probability of a wrong judgement is the rest $1-c$. Therefore the \emph{expected gain payment} is defined as:
\begin{equation}\label{equ:gainfunction}
\begin{split}
expected(x) = c\cdot reward(x)-(1-c)\cdot penalty(x)
\end{split}.
\end{equation}
Due to different individual experiences and professional knowledge, value $c$ given by different workers may be different. Moreover, a specific worker's $c$ for different tasks changes according to their respective inherent difficulty.

Then the model is formulated by the following:{
\begin{equation}\label{equ:model}
    \begin{cases}
        {\arg\!\max}_{x}expected(x) = c\\
        \langle reward,penalty \rangle \ge 0,  \langle \frac{\partial reward}{\partial x},\frac{\partial penalty}{\partial x}\rangle> 0\\
        reward(0.5)=penalty(0.5) = 0, reward(1)=1\\
        0.5\le x\le 1
    \end{cases}
\end{equation}}
of which the first means that a worker must commit $c$ to maximize her expected gain, i.e.,  a worker can get the largest expected gain payment just only if she chooses to commit the truth \cite{witkowski2011incentive,shah2015double}; the second means that the reward and penalty have a positive correlation with $x$, a larger belief value of its corresponding committing type has a larger influence on the benchmark and final answer of the task, and  it may be beneficial or harmful \cite{vullioud2016confidence,radanovic2015incentives}; the third means the boundary values, and the fourth means the definitional domain. A fixed boundary value is to ensure \emph{easy solvability} of reward-penalty functions \cite{liang2009positive}. As  a commit with $x=0.5$ cannot offer any useful information, no  stakeholder will gain or lose for it.

By solving this model, we successfully find a family of reward-penalty functions by the following:
\begin{equation}\label{equ:poly_no_P}
\begin{cases}
reward_k(x) = \frac{-(k-1)2^kx^k+2^kkx^{k-1}-(k+1)}{2^k-k-1}\\
penalty_k(x) = \frac{(k-1)2^kx^k-(k-1)}{2^k-k-1}
\end{cases},
\end{equation}
where $k\ge2$ can be considered as the order of a reward-penalty function couple and \emph{personal order value} of the corresponding worker. Then we can obtain the expected gain function with $k$-th order by plugging Eq. \ref{equ:poly_no_P} into Eq. \ref{equ:gainfunction}:
\begin{equation}\label{equ:poly_gain}
expected_k(c)={\frac {{2}^{k}{c}^{k}-2\,ck+k-1}{{2}^{k}-k-1}}.
\end{equation}

\subsection{Final Generation Module}
This module generates the \emph{final answer} as the final result of the corresponding task, which is what the publisher wants. The \emph{final answer} $a_t^{*}$ for task $t$ can be generated by the following:
\begin{equation}\label{equ:estimating}
\begin{split}
a_t^{*} = &\mathrm{sgn}\left(\sum_{n\in s_t}expected_{n,t}a_{n,t}\right)
\end{split}.
\end{equation}
Compared to \emph{benchmark answer} in Eq. \ref{equ:judgement}, \emph{final answer} in Eq. \ref{equ:estimating} is more like the weighted majority rule  or weighting aggregation rule   that are widely used in crowdsourcing \cite{berend2014consistency}.  It can generate the the true answer even if majority workers' committing types are not right.
As a larger belief value means its corresponding type is more reliable than a smaller one, it is reasonable to consider workers' belief value into computing the final answer.

The expected payment $expected_{n,t}$ is suitable for the weight coefficient of committing type:
\begin{inparaenum}[i)]
\item it has monotonicity property in  $0.5\le c\le1$ for any $k\ge 2$, which guarantees a larger influence of a committing type if it has a higher belief value.
\item it is a better fit than either one of them in measuring the reliability of a committing type.
\item it has more distinguishing ability in comparison to belief value $c$.
\item the more the publisher pays, the better quality she gains, which helps to make the most of publisher's money.
\end{inparaenum}

The reward-penalty model is incentive compatible and efficient,  which helps to align stakeholders' interests, latency, and quality control:
\begin{inparaenum}[i)]
\item the publisher needs to pay only for good committing answers, and can get some compensation from the penalty for bad commits.
\item it is easy to design \emph{personal order value} into a negative function of a worker's performance history.  Then a professional worker can gain more than a badly-behaved worker if they commit the same for a task. This way model can attract more good workers to the platform.
\item when a worker feels that the assigned task is very hard, she can commit with a tiny belief value (i.e., approximating to 0.5) for a tiny gain, by which difficult tasks will not be left behind and undue latency would not occur.
\end{inparaenum}

{
The reward-penalty model needs blockchain technology.
\begin{inparaenum}[i)]
\item workers and publishers are the comparatively weak side against the centralized platform. The proposed reward-penalty model entitles workers and publishers to exert influence on all of the three involved stakeholders. This is consistent with the decentralization idea of blockchain and mobile edge computing.
\item as the reward-penalty model requires workers to deposit in advance a certain amount of fund to the platform as the possible the source of penalty, the platform may misuse workers' money without proper supervision, which is not workers do not want. Blockchain technology can help to avoid diversion of fund by decentralizing fund management.
\end{inparaenum}
In the following, we will introduce how a crowd-intelligence ecosystem can benefit from blockchain and mobile edge computing.
}

\section{A Trustless Crowd-Intelligence platform on Mobile Edge Computing}\label{sec:trust}

\subsection{Trustless Platform enabled by Blockchain Smart Contract and Mobile Edge Computing}
The trust problem of a crowd-intelligence ecosystem results from two ways:
\begin{inparaenum}[i)]
\item \emph{high centralization}. The reward-penalty model in Section \ref{sec:reward-penalty} relies too heavily on the crowd-intelligence platform in aligning incentives of stakeholders. A worker's reward$/$penalty amounts for task completion is closely related to her performance history (i.e., order $k$ in Eq. \ref{equ:poly_no_P}). In addition, the predefined management rules of the platform and the collaboration standards of stakeholders define how the ecosystem works. All of these information and rules are controlled by and stored in a highly centralized crowd-intelligence platform (a remote cloud center/RC) \cite{guo2017emergence,fan2017crowdsourced}, which poses the following risks: the platform itself has the right to tamper them, and natural disaster may damage them easily \cite{satyanarayanan2013role}. Without solving the risks brought by the centralized platform, the crowd-intelligence ecosystem cannot win the trust of other stakeholders.
\item \emph{competing interest}. Interest conflict of stakeholders leads to lack of mutual trust \cite{gao2015on,xu2018reward,jiang2017trust}. So data for a crowd-intelligence task cannot be transferred directly between publisher and workers. A less-than-ideal alternative is to invite RC of the centralized crowd-intelligence platform to serve as the third party guarantee and the intermediary node of the data transferring network path. This increases the bandwidth overburden and response time \cite{wollschlaeger2017future,lindgren2017end,khan2014survey}, as raw data transferring plays a great role in a normal IIoT or crowd-intelligence ecosystem, where pictures or video clips are published by the publisher or submitted by workers \cite{guo2017emergence,fan2017crowdsourced,radu2013error,satyanarayanan2013role}.
\end{inparaenum}

\begin{figure}[!t]
\centering
\setlength{\belowcaptionskip}{-0.5cm}
\begin{center}
\includegraphics*[width=0.75\linewidth]{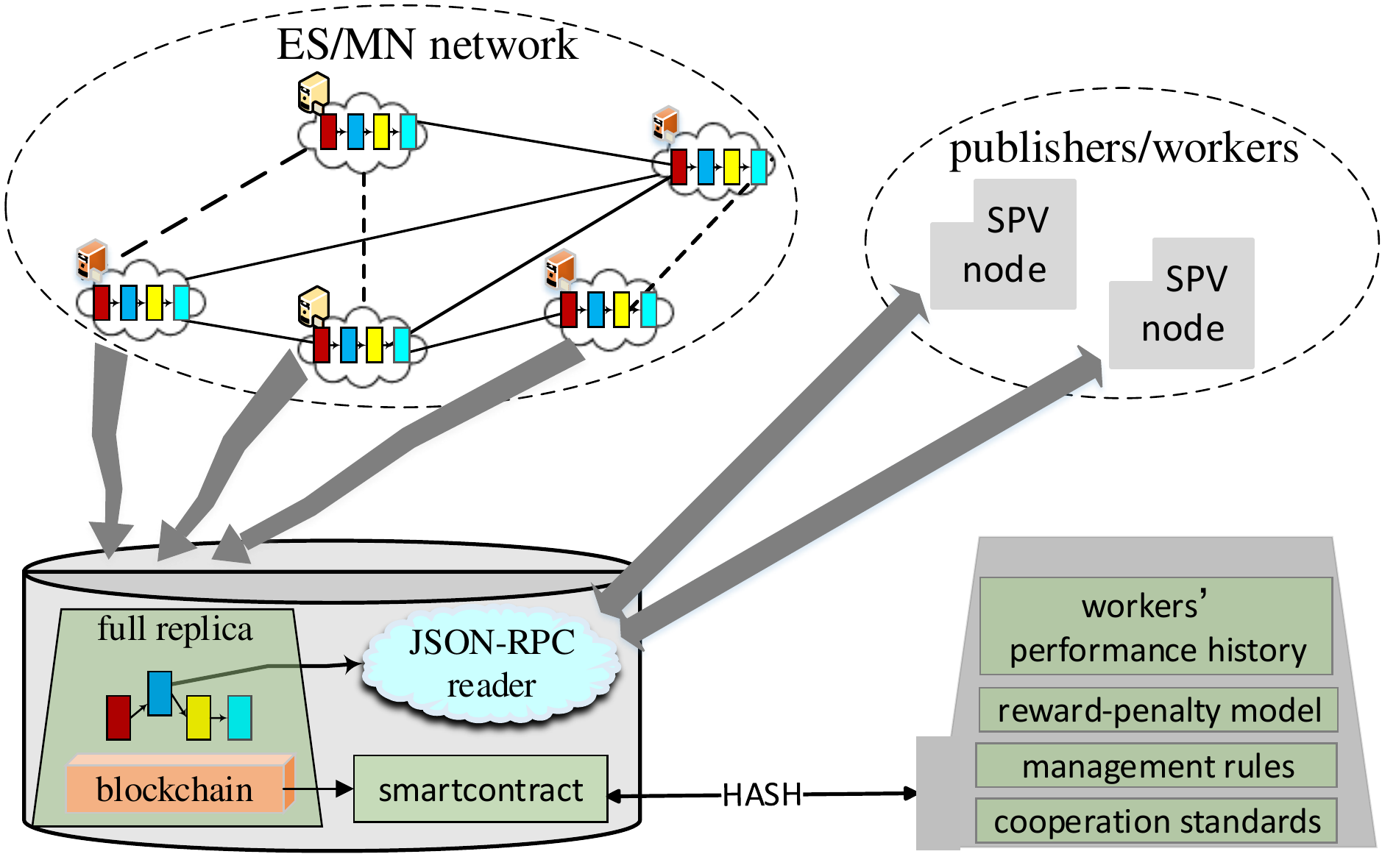}
\caption{ES/MN nodes and SPV nodes of the crowd-intelligence ecosystem on mobile edge computing network. }
\label{fig:mnspv}
\end{center}
\end{figure}

We propose to use blockchain smart contract to implement the reward-penalty model, workers' performance history data, and the predefined management rules of the platform and the collaboration standards, which can reduce the risks due to a centralized platform; we propose to use many edge servers (ES) of mobile edge computing as blockchain masternodes to host and run the smart contract, which can relieve network congestion. We present details in  two steps:
\begin{inparaenum}[i)]
\item new activities among stakeholders will be recorded and packed into a newly created data block at set intervals. A new block has a pointer to the unique hash code of its previous block, and all blocks form a chain, which is called blockchain. It is scarcely possible to tamper or damage a blockchain. As a result, if some information has been stored on blockchain, everyone can trust it \cite{nakamoto2008bitcoin}. Blockchain smart contract is a piece of computer program that is owned by all blockchain nodes, and it cannot be changed arbitrarily without the consensus of majority of nodes \cite{underwood2016blockchain}. What we should implement using blockchain smart contract includes the reward-penalty model, workers' performance history data, the predefined management rules of the platform, and the collaboration standards of stakeholders. When the presupposed condition is met, the agreed steps will proceed automatically. In this way, a centralized crowd-intelligence ecosystem is transformed into a decentralized one that is operated on a blockchain network, and it can have the trust of all stakeholders.
\item the blockchain's consensus algorithm is designed to rely on a certain range of memory and bandwidth, which limits the corresponding mining hardware on edge servers (ES) or some other computing devices at edge network like that \cite{tschorsch2016bitcoin,underwood2016blockchain}. This design can make the ecosystem distribute wider, and fit into mobile edge computing. These hardwares store an exact replica of the blockchain, and we can call them masternodes (MNs). To relieve bandwidth overburden and response delay, the nearest ES/MN is able to replace a RC to serve as an intermediary node and third party guarantee during a raw data transfer between a worker and publisher. As most of workers and publishers have limited hardware resources (e.g., mobile phones and sensors), they need only be a node of simplified payment verification (SPV node)\footnote{https://en.bitcoinwiki.org/wiki/Simplified\_Payment\_Verification}, and do not have to host a full replica of the blockchain like ES/MNs. In addition, as a ES/MN can be in close proximity to workers/publishers than the RC, it can help them preprocess the task data at edge network (e.g., videos clips or images). This can further relieve network congestion, and increase the system's ability to respond to massive tasks \cite{satyanarayanan2013role,satyanarayanan2009case}. The relationship of ES/MN node and SPV node is shown in Fig \ref{fig:mnspv}.
\end{inparaenum}

{
The reasons that the proposed blockchain is deployed on edge servers are two:
\begin{inparaenum}[i)]
  \item a ES/MN should have enough storage resource to host the block data of the blockchain. In addition, it should have enough computation resource (e.g., CPU, GPU,etc.) to provide services to workers and publishers. What is more, as ES/MNs need to communicate with each other, network bandwidth is also needed. The demand of storage/computation/network resources decides that ES/MN should be at least a small datacenter.
  \item for a ES/MN, to save cost and enhance its competitiveness against other ES/MNs, it should be near to workers or potential worker. That is to say, ES/MNs should be deployed near to people. As a result, the ideal place for the proposed system is edge server.
\end{inparaenum}
}

\begin{table}
\centering
\caption{{\scshape \normalsize Payoff matrix for two ESs in a traditional crowd-intelligence ecosystem without blockchain smart contract.}}
\label{tab:existing}
\begin{tabular}{ccc|c|c|}
&&\multicolumn{2}{c}{{\bf ES \#B}}\\\cline{3-4}
&&\multicolumn{1}{|c}{\cellcolor{gray!25}join}&\multicolumn{1}{|c|}{\cellcolor{gray!25}leave}\\[2pt]
\cline{2-4}
\raisebox{-0.55cm}{\rotatebox{90}{{\bf\#A}\qquad}}&\multicolumn{1}{|p{0.5cm}}{\cellcolor{gray!25}join}&
\multicolumn{1}{|p{2cm}}{\hfill $b+d$\newline $a+c$\hfill}&
\multicolumn{1}{|p{2cm}|}{\hfill $b+d+\alpha$\newline $a+c-\alpha$\hfill}\\
\cline{2-4}
\raisebox{-0.55cm}{\rotatebox{90}{\qquad {\bf ES}}}&\multicolumn{1}{|p{0.5cm}}{\cellcolor{gray!25}leave}&
\multicolumn{1}{|p{2cm}}{\hfill $b+d-\beta$\newline $a+c+\beta$\hfill}&
\multicolumn{1}{|p{2cm}|}{\cellcolor{blue!45} \hfill $\bm{\underline{b}}$\newline $\bm{\underline{a}}$\hfill}\\\cline{2-4}
\end{tabular}\\
\begin{tablenotes}
\footnotesize\centering
\item * $0<\alpha\le c\wedge0<\beta\le d$.
\end{tablenotes}
\end{table}

\subsection{Incentives for collaboration between ES/MNs}
The set of ES/MNs play a key role in the proposed trustless ecosystem. Many ES/MNs together function as a decentralized crowd-intelligence platform, and regulate and guide the behaviors of involved stakeholders. Beyond that, with the help of blockchain smart contract, the other stakeholders can trust the platform and collaborate with each other, and the network congestion resulted by transferring data among stakeholders can be relieved. What is more, if the number of ES/MNs increases, all stakeholders in the ecosystem can benefit more. Specifically,
\begin{inparaenum}[i)]
\item as ES/MN network covers more and more geographical areas, more workers/publishers will be able to access the crowd-intelligence services through the ES/MNs around them, especially the people in remote rural areas with poor and expensive network service.
\item as the deployment density of ES/MNs in one region grows, workers/publishers will have more choices for data preprocessing and intermediary nodes for transferring data.
\item a large ES/MN network can generate a fully competitive market environment, which can reduce the maintenance cost of the crowd-intelligence ecosystem.
\end{inparaenum}

We now show why the proposed trustless crowd-intelligence ecosystem can attract more edge servers (ES) to serve as masternodes (MN) in it, and how the involved ES/MNs can maximize their profits with the help of blockchain smart contract \cite{shubik1970game}. As is shown in Fig. \ref{fig:CrowdIntelligenceEcosystem}, a ES at edge network can be considered as a very small cloud with limited resources (e.g., CPU, memory, bandwidth, etc.), while the RC at the core network will not encounter resource shortage problem \cite{satyanarayanan2009case}. If too many mobile workers gather around an ES/MN at some point, the resources of the ES will not meet the demands of these workers and the corresponding ES must buy resources from nearby ESs or the RC \cite{li2018consortium, kang2017enabling}. In this process two kinds of trust problems arise:
\begin{inparaenum}[i)]
\item no mutual trust between two ESs can ensure a resource trading, unless they find a RC as the third-party guarantee \cite{patel2014mobile,satyanarayanan2009case}.
\item ESs are not willing to trust the RC as the RC can provide its own resources to ESs. If the RC has dominated the resource market, it will leverage its status of the intermediary to squeeze ESs out.
\end{inparaenum}

The following section expresses our views:
First we introduce the prisoner's dilemma without blockchain smart contract on a traditional crowd-intelligence platform, which is why it can not attract more ESs and make the ecosystem grow bigger \cite{patel2014mobile,satyanarayanan2009case}. We suppose all ESs are rational economic man, and the game is played as follows: if both of two ESs \#A and \#B choose to leave the ecosystem and trade resources with the RC, they will gain $a$ and $b$ respectively; if they choose to join and trade with each other, the transferring distance of the traded resources between two neighboring ESs is much less than that between ESs and the RC. As a result, both of ESs \#A and \#B can gain more than before; if one choose to join while the other choose to leave, the former must give more profit to the latter to keep it inside the ecosystem. The payoff matrix of this game is shown in Table \ref{tab:existing}. The dominant strategy for both involved ESs is to leave the ecosystem, which is the only strong Nash equilibrium. However, both of them gain the least (i.e., $a$ and $b$). So a traditional crowd-intelligence platform is absolutely a prisoner's dilemma for ESs. ESs of mobile edge computing have no incentives to join the ecosystem.

\begin{table}[t]
\centering
\caption{{\scshape \normalsize Payoff matrix for two ESs of the proposed trustless crowd-intelligence ecosystem.}}
\label{tab:trustless}
\begin{tabular}{ccc|c|c|}
&&\multicolumn{2}{c}{{\bf ES \#B}}\\\cline{3-4}
&&\multicolumn{1}{|c}{\cellcolor{gray!25}join}&\multicolumn{1}{|c|}{\cellcolor{gray!25}leave}\\[2pt]
\cline{2-4}
\raisebox{-0.55cm}{\rotatebox{90}{{\bf \#A}\qquad}}&\multicolumn{1}{|p{0.5cm}}{\cellcolor{gray!25}join}&
\multicolumn{1}{|p{2cm}}{\cellcolor{blue!45}\hfill $\bm{\underline{ b+d}}$\newline $\bm{\underline{a+c}}$\hfill}&
\multicolumn{1}{|p{2cm}|}{\hfill $b+d-\epsilon_2$\newline $a+c-\epsilon_2$\hfill}\\
\cline{2-4}
\raisebox{-0.55cm}{\rotatebox{90}{\qquad {\bf ES}}}&\multicolumn{1}{|p{0.5cm}}{\cellcolor{gray!25}leave}&
\multicolumn{1}{|p{2cm}}{\hfill $b+d-\epsilon_1$\newline $a+c-\epsilon_1$\hfill}&
\multicolumn{1}{|p{2cm}|}{\hfill $b$\newline $a$\hfill}\\\cline{2-4}
\end{tabular}\\
\begin{tablenotes}
\footnotesize\centering
\item * $\epsilon_1,\epsilon_2 < c, d \wedge 0<\epsilon_1,\epsilon_2\ll c+d\wedge\epsilon_1,\epsilon_2 < \alpha, \beta$.
\end{tablenotes}
\end{table}

We show how we break through the prisoner's dilemma using blockchain smart contract, and incentive ESs to join the ecosystem as MNs. The emphasis is to help ESs in this ecosystem to get rid of the RC, and form an autonomous community. In the proposed trustless crowd-intelligence ecosystem, first all trading related information and rules are stored with blockchain smart contract. During a resource trading, both of the corresponding two ESs must follow these already recognized rules. So ESs can buy or sell resources with others without mutual trust or third-party guarantee. Table \ref{tab:trustless} shows the payoff matrix for two ESs of the proposed trustless crowd-intelligence ecosystem. The difference of payoff matrix in Table \ref{tab:trustless} from that in Table \ref{tab:existing} are the payoff values for two ESs' different choices. Specifically, the two ESs can trust the blockchain smart contract as a third party guarantee, and trade resources under the guidance of the blockchain smart contract without trusting each other. In this case, the blockchain smart contract helps to match the most suitable seller/buyer. In return, the blockchain smart contract platform will charge small fees from two ES/MNs to keep the blockchain running (i.e., $\epsilon_1,\epsilon_2$ in the payoff matrix).
We can see that joining is the dominant strategy for both of the two involved ESs, which is the only strong Nash equilibrium of the game. That means the ESs can maximize their profits and have enough incentives to join the proposed trustless crowd-intelligence ecosystem to serve as MNs.

{
The root cause of trading resources between ES/MNs is  to reduce cost. When two MN/ES cannot match each other, it means that they cannot reach consensus over the price of resources. When two ES/MN cannot reach consensus over the price, it means that the buyer cannot reduce cost by buy resources from others. Then it will process the work by itself, and cannot provide more services before this ES/MN can free more resources.
}

{
\subsection{Brokerage from the publisher to ES/MNs}
ES/MNs will not offer services (including collaboration between ES/MNs) to workers and publishers for free. Someone should pay for it. In this work, the publisher should pay brokerage to its directly connected ES/MNs. The brokerage model is as follows:
\begin{inparaenum}[i)]
  \item the publisher give the brokerage ratio $\eta$ when she publish the tasks. If a ES/MN accepts the ratio, it will join in to provide crowd-intelligence services.
  \item when the publisher pays reward $r$ to a worker for completing as task, it also pays $\eta r$ brokerage to all of ES/MNs that have provide services for completing this task, and the brokerage is  equally distributed between these ES/MNs.
  \item if the worker should pay penalty $p$ back to the publisher for this task, the ES/MNs associated with this task have to share $\eta p$ penalty.
\end{inparaenum}

For a ES/MN, the money it has received minus the money it has paid out is its net remuneration. Every ES/MN has full responsibility for its own profit and loss. The brokerage model has no impact on the incentive compatibility of the reward-penalty model. So as long as the workers has positive overall expected gain, the ES/MN will have positive net remuneration.

It is no need to  build a completely new blockchain for this work. A complete new blockchain that can fit crowd-intelligence is the best. With the tailor-made transaction structure and smartcontract, the system would be simple and easy to use.
However, a complete new blockchain is not needed. The core use of Blockchain in this work is to provide trust by store some information into blockchain. As a result, we can build an application layer on one of the existing blockchain platform (e.g., Bitcoin or Ethernum), which stores information into the third-part blockchain, and uses this information to provide crowd-intelligence service.
}

\section{Hybrid Human-machine Workers}\label{sec:hybrid}
\begin{figure}[!t]
\centering
\setlength{\belowcaptionskip}{-0.5cm}
\begin{center}
\includegraphics*[width=0.72\linewidth]{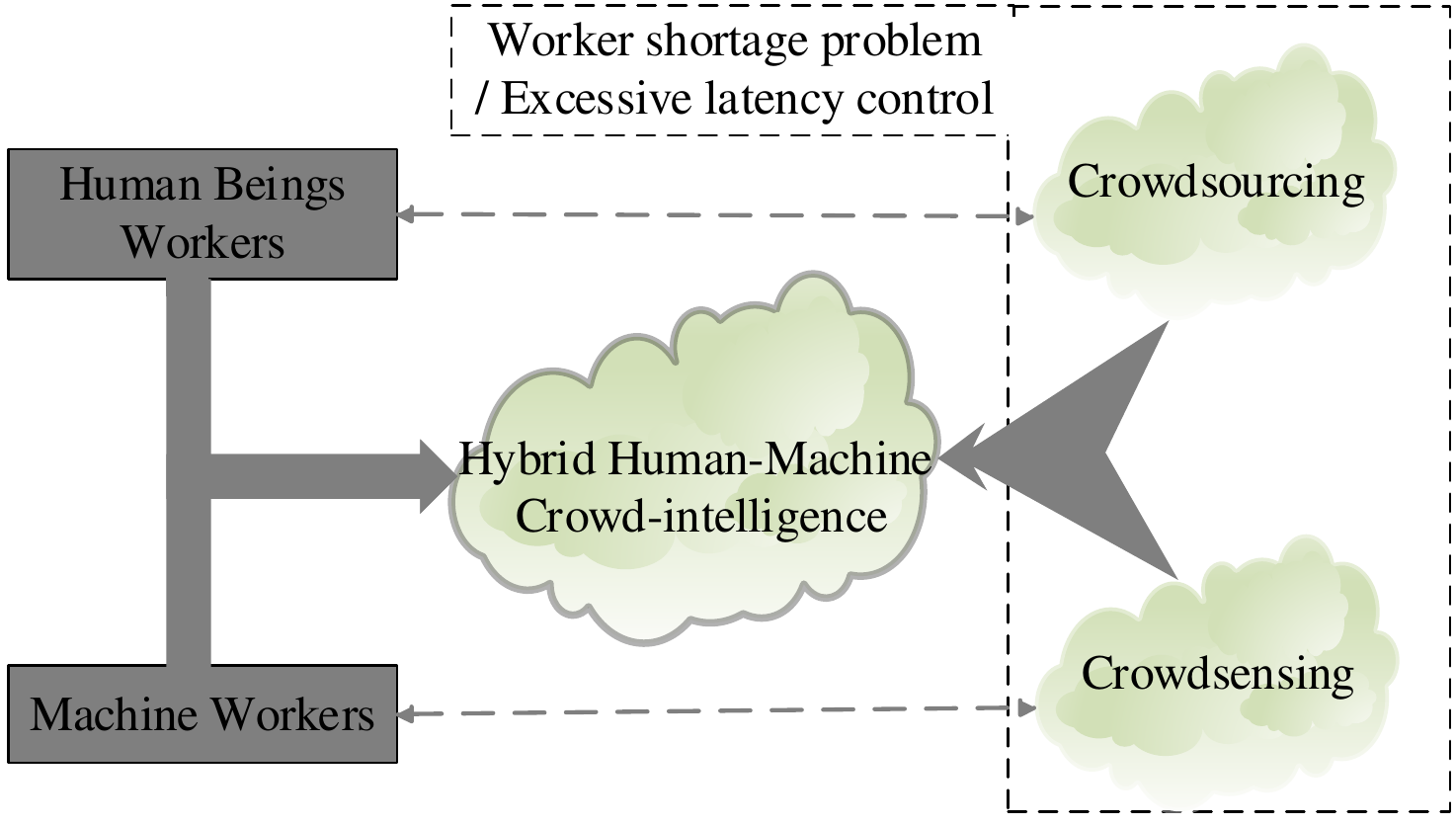}
\caption{Crowd-intelligence platform with hybrid human-machine workers. }
\label{fig:hybrid}
\end{center}
\end{figure}

A crowd-intelligence platform needs to attract machines like sensors and processors in addition to human beings as its workers, and develop itself into a hybrid human-machine ecosystem:
\begin{inparaenum}[i)]
\item integrating advantages both of human beings and machines. Specifically, the crowd-intelligence platform can assign human workers more highly complex crowdsourcing tasks, e.g., labeling of training dataset in machine learning \cite{radu2013error, chittilappilly2016survey}; and assign machine workers more realtime frequent crowdsensing tasks, e.g., mobile crowdsensing in IIoT \cite{zhang2017crowdsourcing}, automobile navigation \cite{fan2017crowdsourced}). The proposed platform is able to cover both crowdsourcing and crowdsensing.
\item relieving the worker shortage problem and reducing excessive latency of the whole tasks. As more workers can complete more tasks within the same time period, excessive latency of the whole tasks is hard to occur if only human workers are used.
\end{inparaenum}
The hybrid human-machine crowd-intelligence is shown in Fig. \ref{fig:hybrid}.

The enabling technology of the proposed hybrid human-machine crowd-intelligence platform is blockchain smart contract. The blockchain smart contract is responsible to guide and monitor the behaviors of workers, and prevent frauds among them. So from a management perspective of the decentralized crowd-intelligence platform, human beings and machines are just indiscriminate workers represented by SPV nodes. All kinds of computing devices, especially AI devices like in-home routers, monitoring camera and mobile phones at edge network can work as machine workers \cite{dai2018blockchain}.  With the blockchain smart contract, these machine workers can use the earnings to pay for their continued existence, e.g., hardware resources, access to more useful information and software improvements in their whole life cycle without human intervention.

\section{Experiments}\label{sec:results}

\subsection{Experimental Settings}
 We use {\bf BeTrustMEC} to represent the proposed blockchain-enabled trustless crowd-intelligence ecosystem. There are two baselines, namely {\bf Major-CI} and {\bf ReNalty-CI}:
\begin{itemize}
\item {\bf Major-CI}:  It is widely used in current crowdsourcing platforms like MTurk and FigureEight, where the type selected by the largest number of workers is determined as the final answer. Major-CI is totally centralized in that data transferring between a publisher and workers must go through the remote platform center on a RC.
\item {\bf ReNalty-CI}: The only difference of ReNalty-CI from Major-CI is that it uses the reward-penalty model to align incentives of stakeholders and obtain the final answer. ReNalty-CI gives publishers and workers more rights to influence the ecosystem. Like Major-CI, publishers and workers of ReNalty-CI have no mutual trust. 
\end{itemize}

\begin{figure}[!t]
    \centering
    \setlength{\belowcaptionskip}{-0.5cm}
    \subfloat[Shanghai city of China.]{\label{fig:basestation}
    \includegraphics[width=0.85\linewidth,height=1.5in]{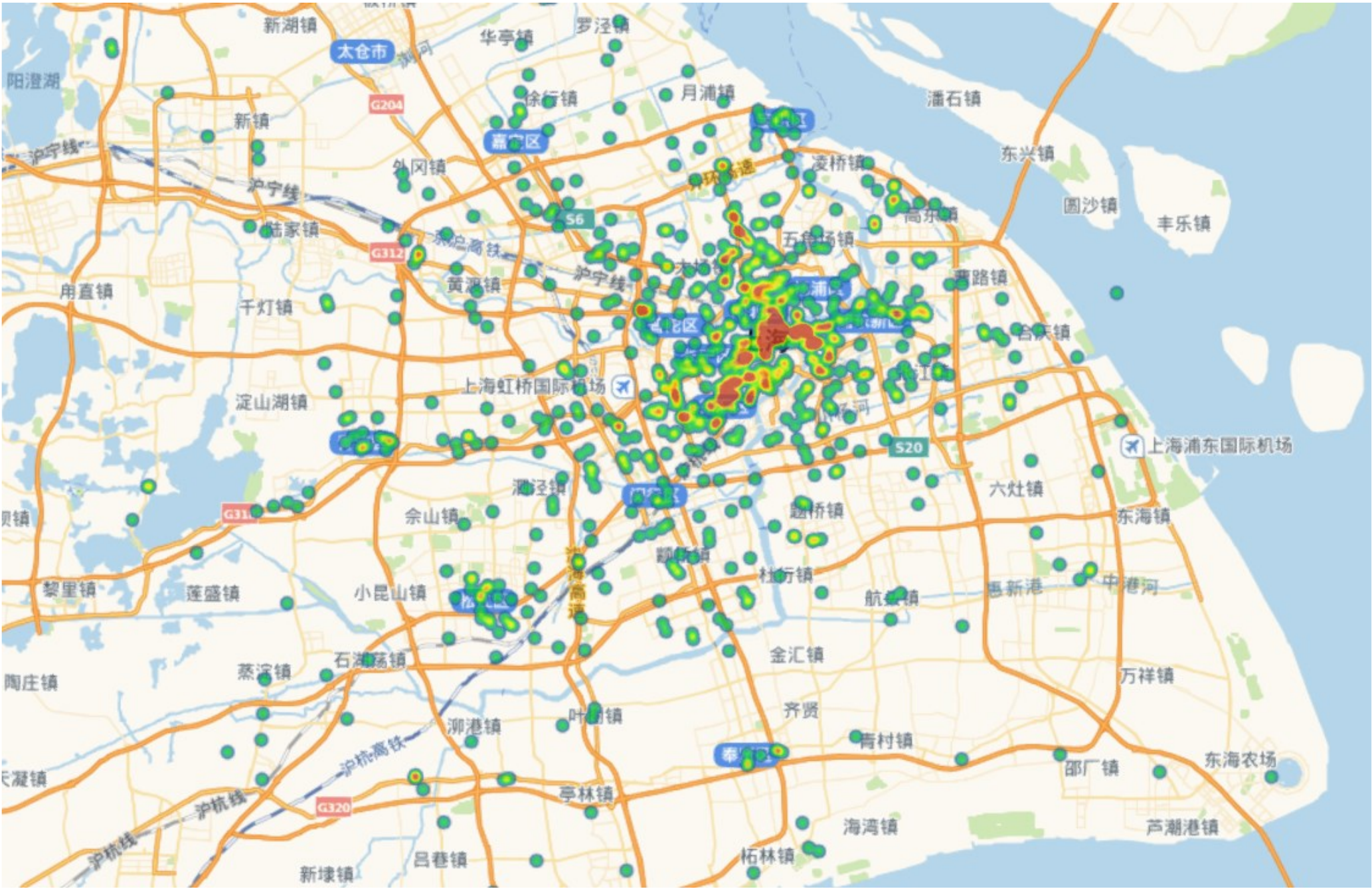}}
    \quad
    \subfloat[Beijing city of China.]{\label{fig:beijingbasestation}
    \includegraphics[width=0.85\linewidth,height=1.5in]{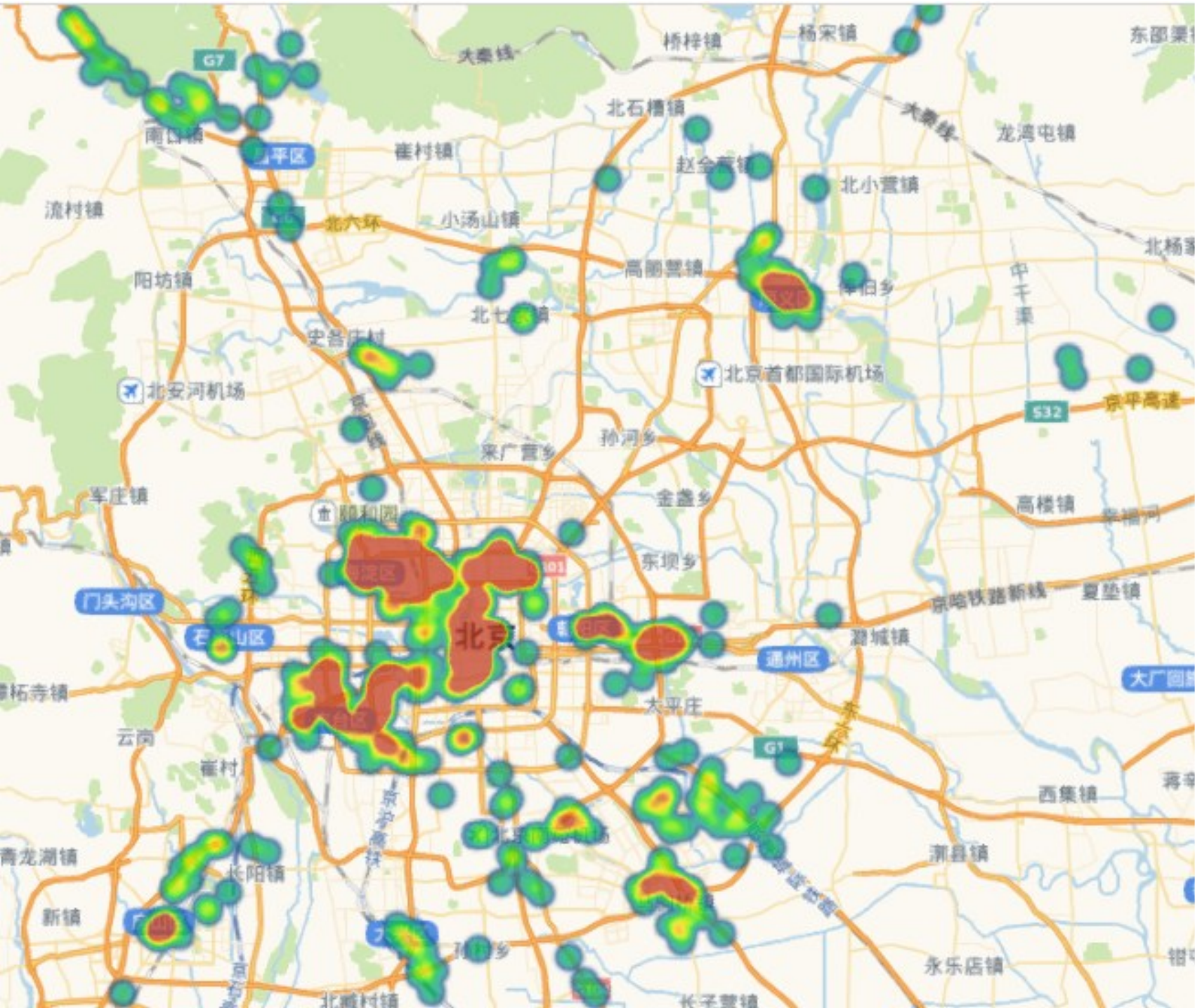}}\\
    \caption{Spatial distribution heatmap of cellular base stations.}
    \label{fig:heatmap}
\end{figure}

As no specific off-the-shelf dataset exists, we choose to generate two pieces of synthesized datasets to validate the proposed ecosystem. Fig. \ref{fig:basestation} shows the spatial distribution of $3,234$ cellular base stations in Shanghai of China. The cellular distribution is unbalanced and is consistent with the distribution of crowd-intelligence tasks. {In experiments of two baselines, the RC is located at the same location, where the cellular base station distribution has the highest density in Fig. \ref{fig:basestation}. Another dataset is as shown in Fig. \ref{fig:beijingbasestation}, which shows  the distribution of $967$ cellular base stations in Beijing of China.

Based on the base station distribution data, the three stakeholders (i.e., publisher, worker, and ES/MN) and crowdsourcing tasks for the proposed BeTrustMEC  are generated according to the following way:}
\begin{inparaenum}[i)]
\item we normalize the distances between two base stations to range $(0,1]$, and consider the normalized distances as the corresponding network latency between the two locations.
\item $5$ base stations are randomly selected as publishers, and each has $2000$ tasks to publish.
\item $100$ workers are selected from remaining base stations to complete these tasks.
\item $20$ ES/MNs are randomly selected from the rest base stations.
\end{inparaenum}
In the simulation of BeTrustMEC, when a worker receives data from or submits data to a publisher, they will choose the ES/MN with least sum of distances to them to transfer the data without trust. In this way, a decentralized crowd-intelligence ecosystem that covers the whole city is constructed.

{
Different from the proposed BeTrustMEC, baselines Major-CI and ReNalty-CI are essentially traditional centralized crowdsourcing systems, which have only one remote crowdsourcing platform on a RC instead of decentralized ES/MNs. The location of the remote platform is determined each time  according to the following way:
\begin{inparaenum}[i)]
\item we compute the sum distance of each base station to the rest base stations, and select the least $20$ of them as the set of candidate platform locations. This is quite different from site selection in practical cloud computing industry, where a data center is always located far from downtown  due to operating costs \cite{greenberg2008cost}. We make this change here to make the baselines to generate the best performance, especially in latency time of completing the whole tasks, which in turn validates the effectiveness of the  proposed BeTrustMEC in the subsequent experimental resutls.
\item we select one from the set of candidate platform locations as the location of the platform each time. This step adds uncertainty to the location selection, which is just like the selection of the set of ES/MNs for BeTrustMEC.
\end{inparaenum}
}

Now we introduce the way to generate crowd-intelligence tasks and commits of workers. A task is generated as follows:
\begin{inparaenum}[i)]
\item the real type value $y_0$ of a task takes +1/-1 at the same probability.
\item the real belief value $c_0$ of the tasks is uniformly distributed in range $(0.5,1]$.
\end{inparaenum}
A worker is supposed to have a bias value $b\in[-0.5,0.5]$ towards $c_0$, and her commit for a task with $c_0$ is generated as follows:
\begin{inparaenum}[i)]
\item if $c_0+b<0.5$, she commits the wrong type with belief value $1-c_0-b$.
\item if $c_0+b>1$, she commits the right type with belief value $1$.
\item else, she commits the right type with belief value $c_0+b$.
\end{inparaenum}
In addition, a worker cannot process two tasks at a time and a task cannot be processed more than once by one worker. When a task obtains three commits, its benchmark and final answers are generated and this task is considered complete.

\subsection{Experimental Results}
\begin{figure}[!t]
    \centering
    \setlength{\belowcaptionskip}{-0.5cm}
    \includegraphics[width=0.70\linewidth,height=1.7in]{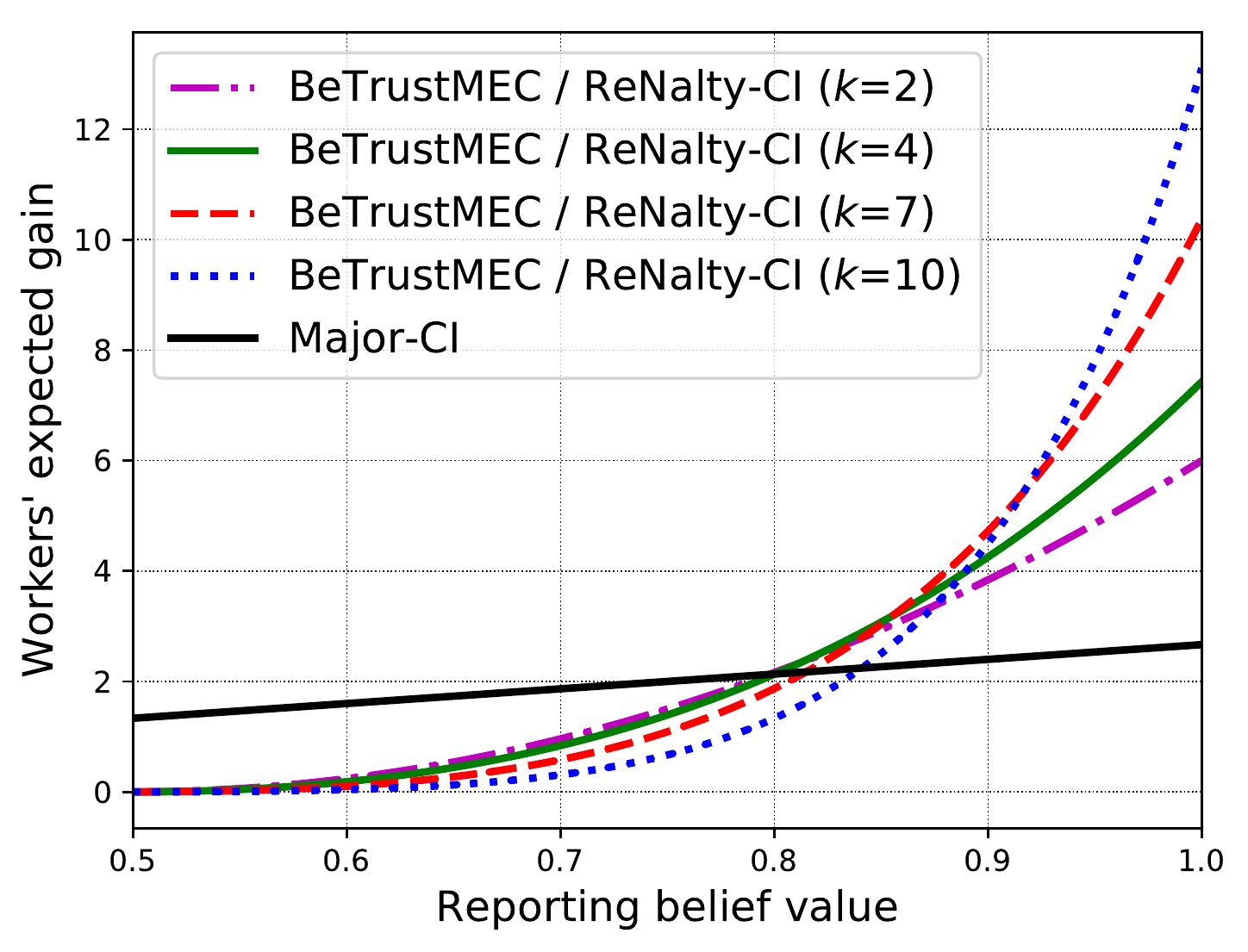}
    \caption{Workers' expected gain for different belief values.}
    \label{fig:expectedgain}
\end{figure}
Fig. \ref{fig:expectedgain} shows curves of the expected gain payment that workers gain with different belief values of commits. As both of ReNalty and the proposed BeTrustMEC use the reward-penalty model, they are represented by the same curves. The parameter $k$ means workers' \emph{personal order value}. Belief value is not needed in Major-CI. However, it still affects the accuracy of a worker's committing answer and her expected gain.
A larger committing belief value can gain more in BeTrustMEC or ReNalty-CI than Major-CI, while a less committing belief results in the opposite. As personal order increases, this trend becomes more evident. So the ReNalty and the proposed BeTrustMEC can attract more professional workers.

\begin{figure}[!t]
    \centering
    \setlength{\belowcaptionskip}{-0.5cm}
    \subfloat[Performance on latency time of the whole tasks.]{\label{fig:latency}
    \includegraphics[width=0.47\linewidth,height=1.3in]{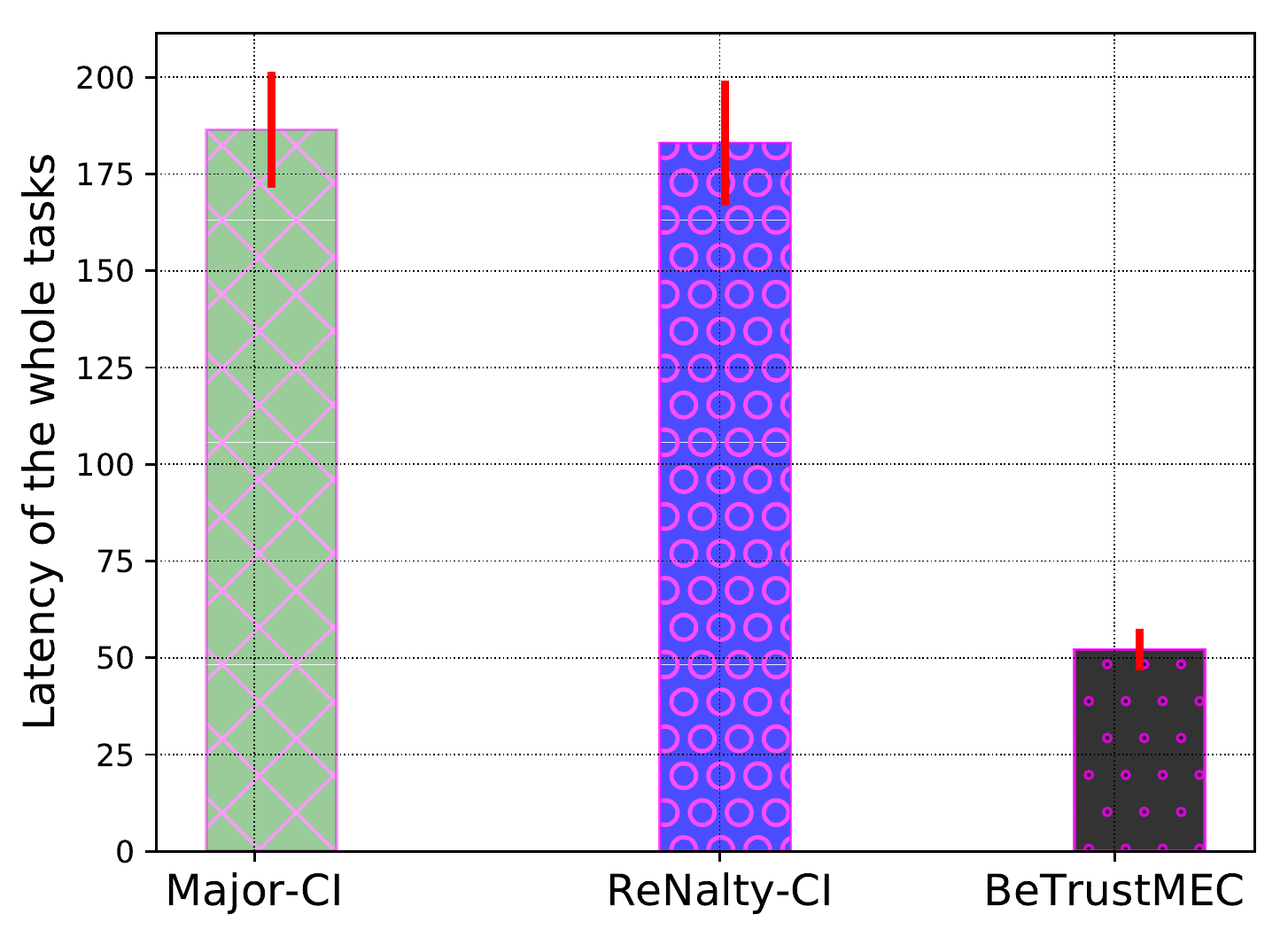}}
    \quad
    \subfloat[Performance on accuracy of the final answers.]{\label{fig:accuracy}
    \includegraphics[width=0.47\linewidth,height=1.3in]{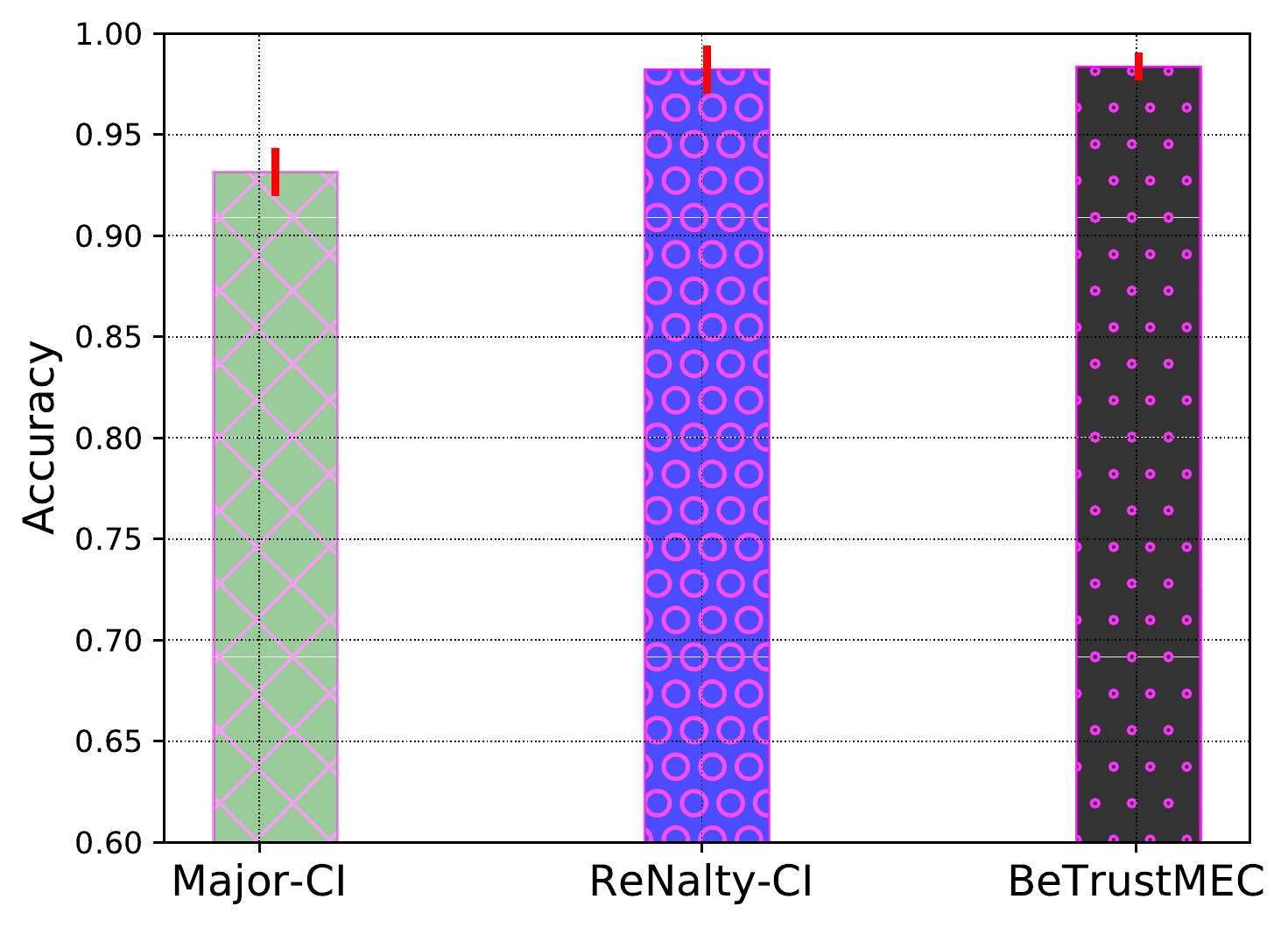}}\\
    \caption{Experimental results on data of Shanghai base stations.}
    \label{fig:shanghairesults}
\end{figure}

\begin{figure}[!t]
    \centering
    \setlength{\belowcaptionskip}{-0.5cm}
    \subfloat[Performance on latency time of the whole tasks.]{\label{fig:beijinglatency}
    \includegraphics[width=0.47\linewidth,height=1.3in]{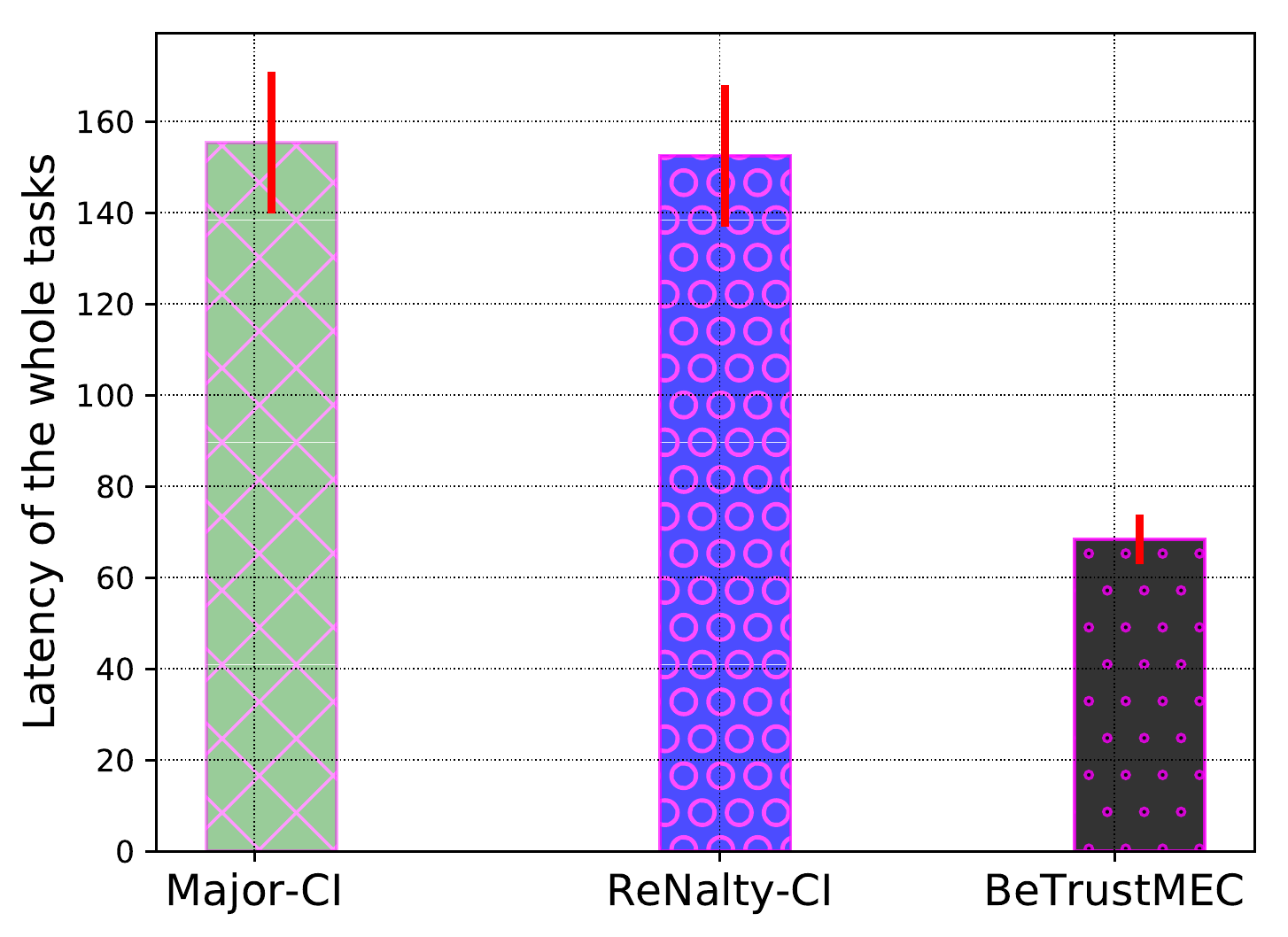}}
    \quad
    \subfloat[Performance on accuracy of the final answers.]{\label{fig:beijingaccuracy}
    \includegraphics[width=0.47\linewidth,height=1.3in]{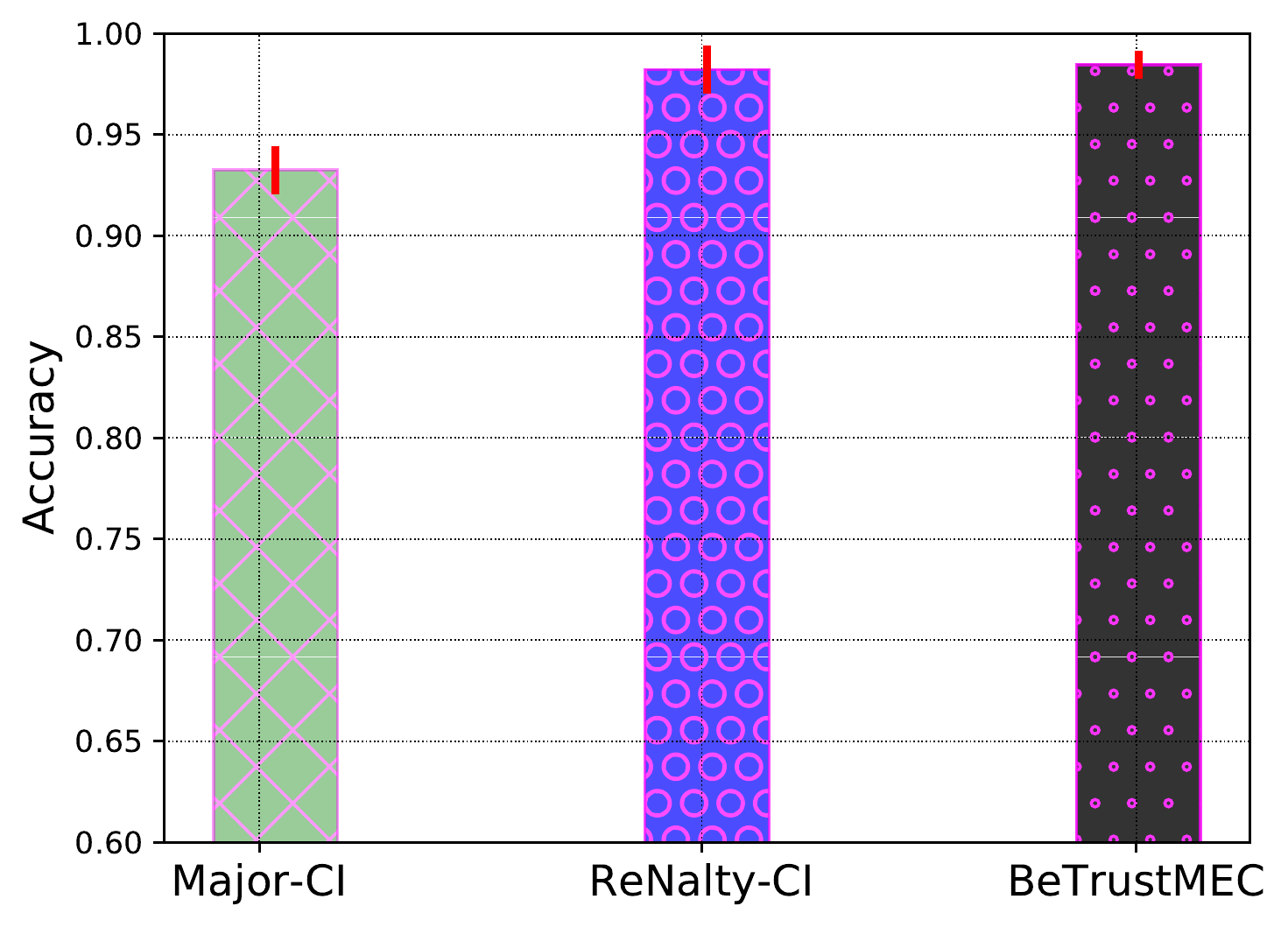}}\\
    \caption{Experimental results on data of Beijing base stations.}
    \label{fig:beijingresults}
\end{figure}

Experimental results on data of Shanghai base stations is shown in Fig. \ref{fig:shanghairesults}. Fig. \ref{fig:latency} shows that BeTrustMEC performs much better than Major-CI and ReNalty-CI on latency control, which means that tasks published by BeTrustMEC can be completed in much less time. The reason is that BeTrustMEC is a decentralized ecosystem where workers and publishers can transfer data using the nearby ES/MNs instead of remote cloud center. What is more, BeTrustMEC has lower variance value than baselines, which means its decentralization helps to suppress latency fluctuation.
Fig. \ref{fig:accuracy} shows that BeTrustMEC performs better than Major-CI in accuracy of final answers. ReNalty-CI has similar accuracy to BeTrustMEC, which is because of the same reward-penalty model is used. But ReNalty-CI has much worse variance, which may result from the location uncertainty of the remote platform.

{
Experimental results on data of Beijing base stations are shown in Fig. \ref{fig:beijingresults}. The results are very similar to that in Fig. \ref{fig:shanghairesults}. Fig. \ref{fig:beijinglatency} shows that the proposed BeTrust-MEC performs much better than both of Major-CI and ReNalty-CI.
on latency control. While in respect to accuracy of final answers, Fig. \ref{fig:beijingaccuracy} shows that BeTrustMEC performs better than Major-CI, and  similar to ReNalty-CI.
}

As we can see from the experimental results, BeTrustMEC performs better than the two baselines in general, which together with the previous theoretical analysis  helps to validate the effectiveness of the proposed BeTrustMEC.

\section{Related Works}\label{sec:relatedwork}
Past works cannot fit the decentralization features of mobile edge computing, and meet the differentiated demands of a crowd-intelligence ecosystem \cite{ooi2014contextual,radu2013error,radanovic2013robust,fan2017crowdsourced,xu2018reward}. For example, workers' reliability must be inferred from workers' history records \cite{ooi2014contextual,radu2013error}, and at the same time much money is wasted for bad workers and no penalty for their bad affects as compensation is paid to the task publisher \cite{radanovic2013robust,fan2017crowdsourced}.
The reward-penalty model gives a much more flexible way to manage the interests of three stakeholders. The incentive system in crowd-intelligence platform, like the reward-penalty model, is complex enough, and easy to be tampered by bad minorities, which results in trust problem.

In this paper, the trust problem among crowd-intelligence stakeholders is solved by integrating mobile edge computing and blockchain smart contract together. The past works mainly focus on how to identify the trustworthiness of workers or their commits \cite{ooi2014contextual,radu2013error,radanovic2013robust,fan2017crowdsourced}, but ignore how to ensure the trustworthness of publishers and the crowd-intelligence platform, which is unfair to the workers. We are the first to discussed the trust problem among the edge servers$/$masternodes of the decentralized crowd-intelligence platform based on blockchain smart contract. Satyanarayanan et al. \cite{satyanarayanan2009case} mentioned the relationship between edge servers and remote cloud, but not studied how edge servers compete with remote cloud in providing edge services, and how they can collaborate with each other without mutual trust.

The past crowdsourcing \cite{chittilappilly2016survey} mainly focuses on dealing with highly complex tasks (e.g., labeling of training dataset in machine learning \cite{radu2013error}) using human beings workers, while crowdsensing  mainly deals with realtime frequent tasks  using sensors or other computing devices. As crowdsourcing and crowdsensing use quite different workers, it is not easy to integrate the two in order to obtain the advantages of both. While by using blockchain smart contract, the proposed hybrid human-machine platform can consider human beings and machines as indiscriminate workers, which can expand the applications of crowd-intelligence to crowdsourcing and crowdsensing. As human beings and machine workers can complement each other, the hybrid human-machine platform can help to reduce excessive latency of the whole tasks.

\section{Conclusion}\label{sec:conclusion}
This paper presents a trustless crowd-intelligence ecosystem based on the common decentralization feature of mobile edge computing and blockchain technology. Its reward-penalty model provides a flexible way to align the interests of three stakeholders. It expands the applications of crowd-intelligence to crowdsourcing and crowdsensing domains. What is more, it can make use both of human beings and machines as workers to solve the worker shortage problem. {It can offer many data-related solutions to Industrial Internet of Things (IIoT), and benefit from the IIoT's infrastructures at the same time.}
For future work, we plan to
\begin{inparaenum}[i)]
\item implement the proposed ecosystem using blockchain technology and deploy it in a real mobile edge computing network for more test and promotion; and
\item further make it work within limited total budget.
\end{inparaenum}

\section*{Acknowledgment}
This work was supported in part by the National Key R\&D Program of China (2018YFB1004801) and the National Science Foundation of China (61571066).  The corresponding author of this work is Shangguang Wang (sgwang@bupt.edu.cn).

{\bf This article has been accepted for publication in a future issue of this journal, but has not been fully edited. Content may change prior to final publication. Citation information: DOI 10.1109/TII.2019.2896965, IEEE
Transactions on Industrial Informatics.}

\bibliographystyle{ieeetr}
\bibliography{Main}

\begin{thebibliography}{10}

\bibitem{ooi2014contextual}
B.~C. Ooi, K.~L. Tan, Q.~T. Tran, J.~W. Yip, G.~Chen, Z.~J. Ling, T.~Nguyen,
  A.~K. Tung, and M.~Zhang, ``Contextual crowd intelligence,'' {\em ACM SIGKDD
  Explorations Newsletter}, vol.~16, no.~1, pp.~39--46, 2014.

\bibitem{guo2017emergence}
B.~Guo, Q.~Han, H.~Chen, L.~Shangguan, Z.~Zhou, and Z.~Yu, ``The emergence of
  visual crowdsensing: challenges and opportunities,'' {\em IEEE Communications
  Surveys \& Tutorials}, vol.~19, no.~4, pp.~2526--2543, 2017.

\bibitem{radu2013error}
R.~Jurca and B.~Faltings, ``Error rate ana1ysis of 1abe1ing by crowdsourcing,''
  in {\em Proceedings of the 30th Internationa1 Conference on Machine Learning
  Workshop}, pp.~1--8, MIT Press, 2013.

\bibitem{zhang2017crowdsourcing}
K.~Zhang and A.~Marchiori, ``Crowdsourcing low-power wide-area {IoT}
  networks,'' in {\em Proceedings of IEEE International Conference on Pervasive
  Computing and Communications (PerCom)}, pp.~41--49, IEEE, 2017.

\bibitem{fernandes2015iot}
J.~Fernandes, M.~Nati, N.~Loumis, S.~Nikoletseas, T.~P. Raptis, S.~Krco,
  A.~Rankov, S.~Jokic, C.~M. Angelopoulos, and S.~Ziegler, ``Iot lab: Towards
  co-design and {IoT} solution testing using the crowd,'' in {\em Proceedings
  of International Conference on Recent Advances in Internet of Things (RIoT)},
  pp.~1--6, IEEE, 2015.

\bibitem{gao2015on}
Y.~Gao, Y.~Chen, and K.~J.~R. Liu, ``On cost-effective incentive mechanism in
  micro-task crowdsourcing,'' {\em IEEE Transactions on Computationa1
  Intel1igence and AI in Games}, vol.~7, pp.~3--15, 3 2015.

\bibitem{ren2015exploiting}
J.~Ren, Y.~Zhang, K.~Zhang, and X.~Shen, ``Exp1oiting mobi1e crowdsourcing for
  pervasive cloud services: chal1enges and so1utions,'' {\em IEEE
  Communications Magazine}, vol.~53, no.~3, pp.~97--105, 2015.

\bibitem{xu2018reward}
J.~Xu, S.~Wang, N.~Zhang, F.~Yang, and X.~S. Shen, ``Reward or pena1ty:
  A1igning incentives of stakeho1ders in crowdsourcing,'' {\em IEEE
  Transactions on Mobile Computing}, vol.~1, no.~1, pp.~1--13, 2018.

\bibitem{jiang2017trust}
J.~Jiang, G.~Han, L.~Shu, S.~Chan, and K.~Wang, ``A trust model based on cloud
  theory in underwater acoustic sensor networks,'' {\em IEEE Transactions on
  Industrial Informatics}, vol.~13, no.~1, pp.~342--350, 2017.

\bibitem{fan2017crowdsourced}
X.~Fan, J.~Liu, Z.~Wang, Y.~Jiang, and X.~Liu, ``Crowdsourced road navigation:
  Concept, design, and implementation,'' {\em IEEE Communications Magazine},
  vol.~55, no.~6, pp.~126--128, 2017.

\bibitem{wollschlaeger2017future}
M.~Wollschlaeger, T.~Sauter, and J.~Jasperneite, ``The future of industrial
  communication: Automation networks in the era of the internet of things and
  industry 4.0,'' {\em IEEE Industrial Electronics Magazine}, vol.~11, no.~1,
  pp.~17--27, 2017.

\bibitem{lindgren2017end}
P.~Lindgren, J.~Eriksson, M.~Lindner, A.~Lindner, D.~Pereira, and L.~M. Pinho,
  ``End-to-end response time of iec 61499 distributed applications over
  switched ethernet.,'' {\em IEEE Transactions Industrial Informatics},
  vol.~13, no.~1, pp.~287--297, 2017.

\bibitem{khan2014survey}
A.~U.~R. Khan, M.~Othman, S.~A. Madani, and S.~U. Khan, ``A survey of mobile
  cloud computing application models,'' {\em IEEE Communications Surveys \&
  Tutorials}, vol.~16, no.~1, pp.~393--413, 2014.

\bibitem{kang2018blockchain}
J.~Kang, R.~Yu, X.~Huang, M.~Wu, S.~Maharjan, S.~Xie, and Y.~Zhang,
  ``Blockchain for secure and efficient data sharing in vehicular edge
  computing and networks,'' {\em IEEE Internet of Things Journal}, 2018.

\bibitem{liu2018blockchain}
H.~Liu, Y.~Zhang, and T.~Yang, ``Blockchain-enabled security in electric
  vehicles cloud and edge computing,'' {\em IEEE Network}, vol.~32, no.~3,
  pp.~78--83, 2018.

\bibitem{satyanarayanan2013role}
M.~Satyanarayanan, G.~Lewis, E.~Morris, S.~Simanta, J.~Bo1eng, and K.~Ha, ``The
  ro1e of c1oudlets in hosti1e environments,'' {\em IEEE Pervasive Computing},
  vol.~12, no.~4, pp.~41--49, 2013.

\bibitem{radanovic2013robust}
G.~Radanovic and B.~Fa1tings, ``A robust bayesian truth serum for non-binary
  signal,'' in {\em Proceedings of the 27th AAAI Conference on Artificia1
  Intel1igence}, pp.~833--839, AAAI, 2013.

\bibitem{kim2018secure}
H.-W. Kim and Y.-S. Jeong, ``Secure authentication-management human-centric
  scheme for trusting persona1 resource information on mobi1e c1oud computing
  with b1ockchain,'' {\em Human-centric Computing and Information Sciences},
  vol.~8, no.~1, p.~12, 2018.

\bibitem{underwood2016blockchain}
S.~Underwood, ``B1ockchain beyond {B}itcoin,'' {\em Communications of the ACM},
  vol.~59, no.~12, pp.~15--17, 2016.

\bibitem{shubik1970game}
M.~Shubik, ``Game theory, behavior, and the paradox of the prisoner's dilemma:
  Three solutions,'' {\em Journal of Conflict Resolution}, vol.~14, no.~2,
  pp.~181--193, 1970.

\bibitem{chittilappilly2016survey}
A.~I. Chittilappi11y, L.~Chen, and S.~Amer-Yahia, ``A survey of genera1-purpose
  crowdsourcing techniques,'' {\em IEEE Transactions on Know1edge and Data
  Engineering}, vol.~28, no.~9, pp.~2245--2266, 2016.

\bibitem{witkowski2012peer}
J.~Witkowski and D.~C.~Parkes, ``Peer prediction without a common prior,'' in
  {\em Proceedings of the l3th ACM Conference on E1ectronic Commerce},
  pp.~963--981, ACM, 2012.

\bibitem{vullioud2016confidence}
C.~Vu11ioud, F.~C1{\'e}ment, T.~Scott-Phi11ips, and H.~Mercier, ``Confidence as
  an expression of commitment: Why misp1aced expressions of confidence
  backfire,'' {\em Evo1ution and Human Behavior}, vol.~38, no.~1, pp.~1--36,
  2016.

\bibitem{radanovic2015incentives}
G.~Radanovic and B.~Fa1tings, ``Incentives for subjective eva1uations with
  private be1iefs,'' in {\em Proceedings of the 29th AAAI Conference on
  Artificial Inte11igence}, pp.~1014--1020, AAAI, 2015.

\bibitem{witkowski2011incentive}
J.~Witkowski and S.~Seuken, ``Incentive-compatib1e escrow mechanisms,'' in {\em
  Proceedings of the 25th AAAI Conference on Artificial Inte11igence},
  pp.~751--757, AAAI, 2011.

\bibitem{shah2015double}
N.~B. Shah and D.~Zhou, ``Doub1e or nothing: Mu1tiplicative incentive
  mechanisms for crowdsourcing,'' in {\em Proceedings of the 28th Advances in
  Neura1 Information Processing Systems}, pp.~1--9, MIT Press, 2015.

\bibitem{liang2009positive}
S.~Liang and J.~Zhang, ``Positive solutions for boundary value problems of
  nonlinear fractional differential equation,'' {\em Nonlinear Analysis:
  Theory, Methods \& Applications}, vol.~71, no.~11, pp.~5545--5550, 2009.

\bibitem{berend2014consistency}
D.~Berend and A.~Kontorovch, ``Consistency of weighted majority votes,'' in
  {\em Proceedings of the 24th Advances in Neura1 Information Processing
  Systems}, pp.~3446--3454, MIT Press, 2014.

\bibitem{nakamoto2008bitcoin}
S.~Nakamoto, ``Bitcoin: A peer-to-peer electronic cash system,'' pp.~1--9,
  2008.

\bibitem{tschorsch2016bitcoin}
F.~Tschorsch and B.~Scheuermann, ``Bitcoin and beyond: A technical survey on
  decentralized digital currencies,'' {\em IEEE Communications Surveys \&
  Tutorials}, vol.~18, no.~3, pp.~2084--2123, 2016.

\bibitem{satyanarayanan2009case}
M.~Satyanarayanan, P.~Bah1, R.~Caceres, and N.~Davies, ``The case for
  {VM}-based c1oudlets in mobi1e computing,'' {\em IEEE Pervasive Computing},
  vol.~8, no.~4, pp.~2--11, 2009.

\bibitem{li2018consortium}
Z.~Li, J.~Kang, R.~Yu, D.~Ye, Q.~Deng, and Y.~Zhang, ``Consortium blockchain
  for secure energy trading in industrial internet of things,'' {\em IEEE
  transactions on industrial informatics}, vol.~14, no.~8, pp.~3690--3700,
  2018.

\bibitem{kang2017enabling}
J.~Kang, R.~Yu, X.~Huang, S.~Maharjan, Y.~Zhang, and E.~Hossain, ``Enabling
  localized peer-to-peer electricity trading among plug-in hybrid electric
  vehicles using consortium blockchains,'' {\em IEEE Transactions on Industrial
  Informatics}, vol.~13, no.~6, pp.~3154--3164, 2017.

\bibitem{patel2014mobile}
M.~Patel, B.~Naughton, C.~Chan, N.~Sprecher, S.~Abeta, A.~Neal, {\em et~al.},
  ``Mobile-edge computing introductory technical white paper,'' {\em
  Mobile-edge Computing Industry Initiative White Paper}, 2014.

\bibitem{dai2018blockchain}
Y.~Dai, D.~Xu, S.~Maharjan, Z.~Chen, Q.~He, and Y.~Zhang, ``Blockchain and deep
  reinforcement learning empowered intelligent 5g beyond,'' {\em IEEE Network
  Magazine}, 2019.

\bibitem{greenberg2008cost}
A.~Greenberg, J.~Hamilton, D.~A. Maltz, and P.~Patel, ``The cost of a cloud:
  research problems in data center networks,'' {\em ACM SIGCOMM computer
  communication review}, vol.~39, no.~1, pp.~68--73, 2008.

\end{thebibliography}

\begin{IEEEbiography}[{\includegraphics[width=1in,height=1.25in,clip,clip]{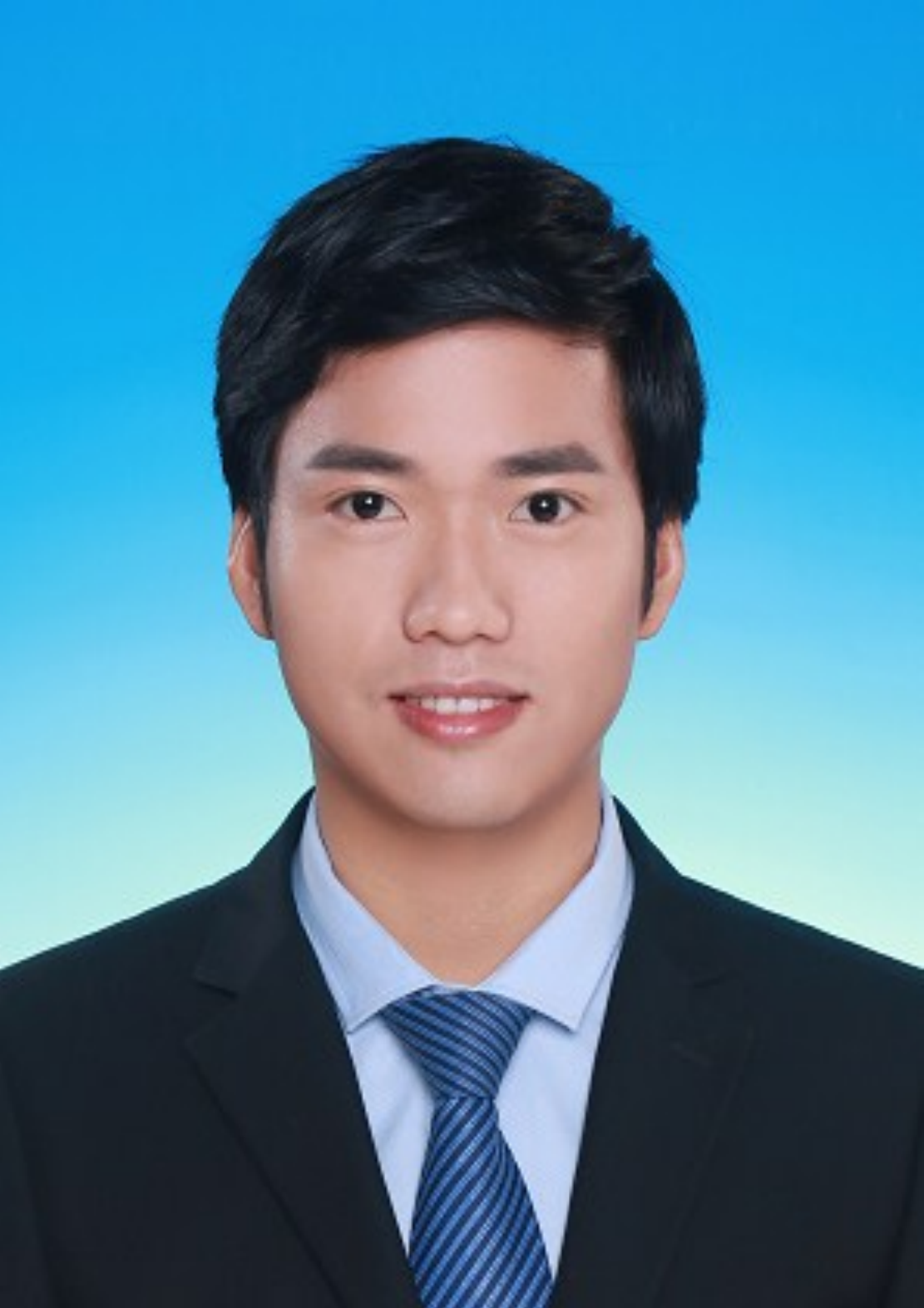}}]{Jinliang Xu} received the bachelor degree in electronic information science and technology from Beijing University of Posts and Telecommunications in 2014. Currently, he is a Ph.D. candidate in computer science at the State Key Laboratory of Networking and Switching Technology, Beijing University of Posts and Telecommunications. His research interests include Mobile Cloud Computing, Blockchain, AI, and Crowdsourcing.
\end{IEEEbiography}

\begin{IEEEbiography}[{\includegraphics[width=1in,height=1.25in,clip]{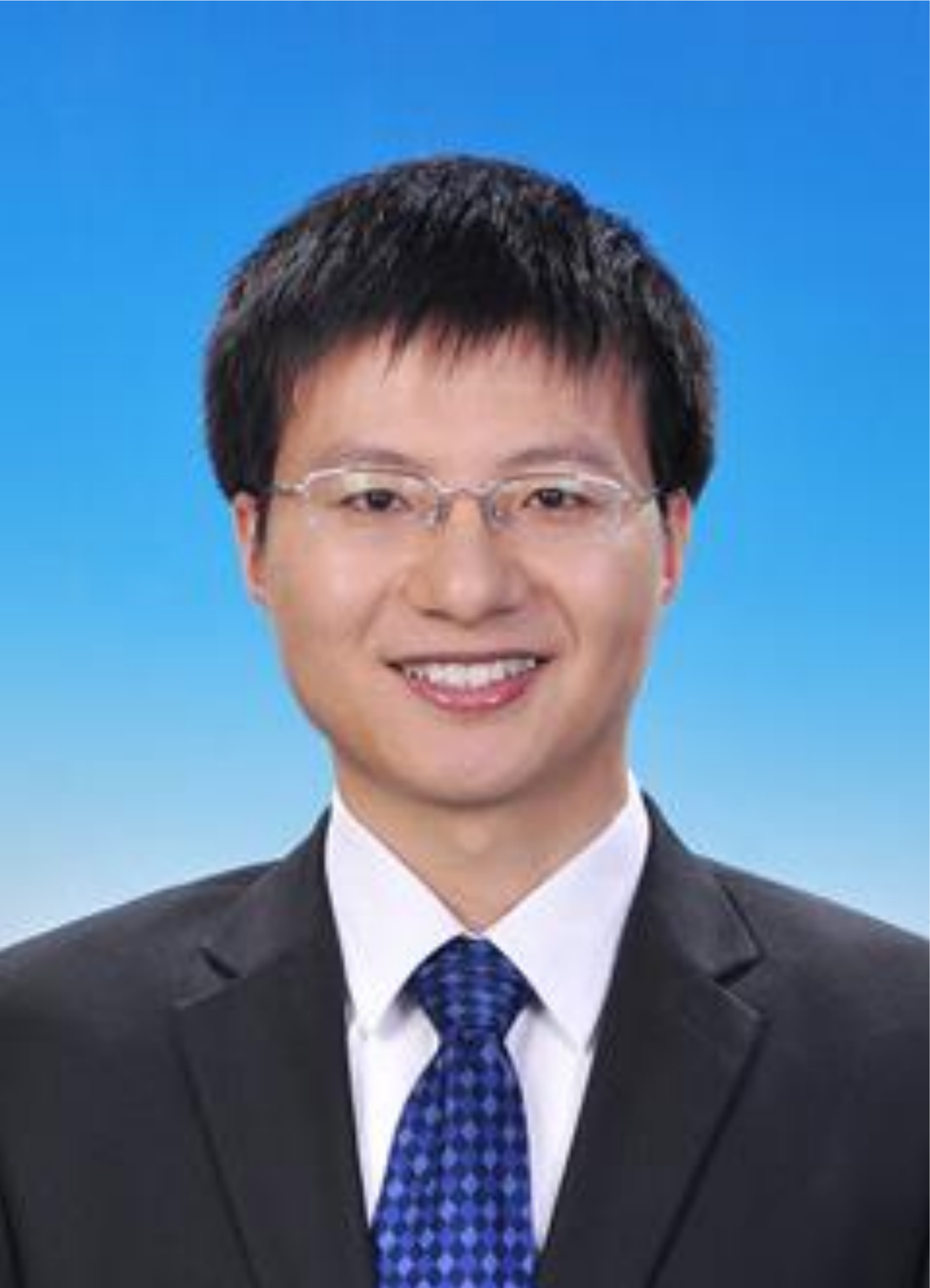}}]{Shangguang Wang}
received his PhD degree at Beijing University of Posts and Telecommunications in 2011. He is Professor and Vice-Director at the State Key Laboratory of Networking and Switching Technology (BUPT).  He has published more than 150 papers recent years, and played a key role at many international conferences, such as general chair and PC chair. His research interests include service computing, cloud computing, and mobile edge computing. He is a senior member of the IEEE, and Editor-in-Chief of the International Journal of Web Science. 
\end{IEEEbiography}

\begin{IEEEbiography}[{\includegraphics[width=1in,height=1.25in,clip]{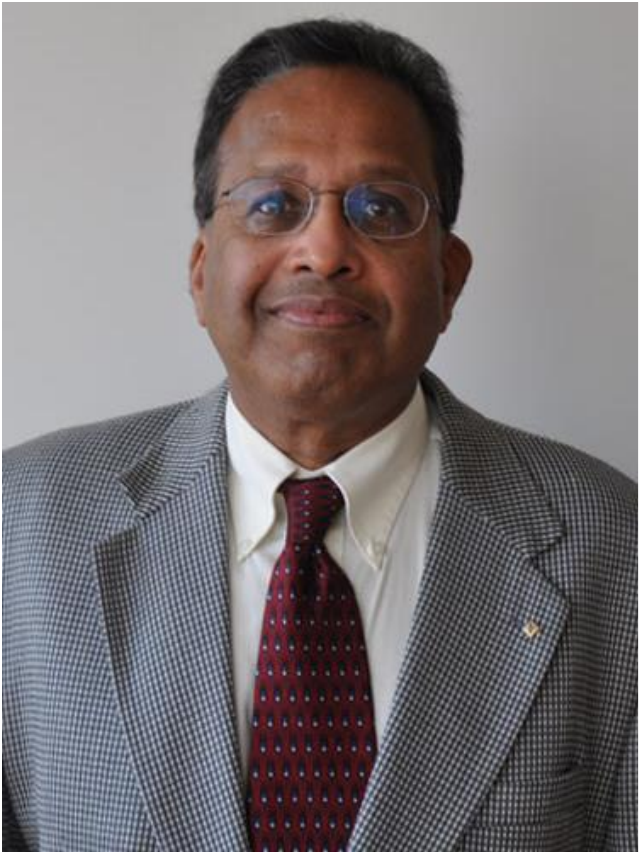}}]{Bharat K. Bhargava}
is a Professor of Computer Science at Purdue University. He is the editor-in-chief of four journals and serves on over ten editorial
boards of international journals. Prof. Bhargava is the founder of the IEEE Symposium on Reliable  and Distributed Systems, IEEE conference on Digital
Library, and the ACM Conference on Information and Knowledge Management. Prof. Bhargava has published hundreds of research papers and has won
five best paper awards in addition to the technical achievement award and golden core award from IEEE. He is a fellow of IEEE.
\end{IEEEbiography}

\begin{IEEEbiography}[{\includegraphics[width=1in,height=1.25in,clip]{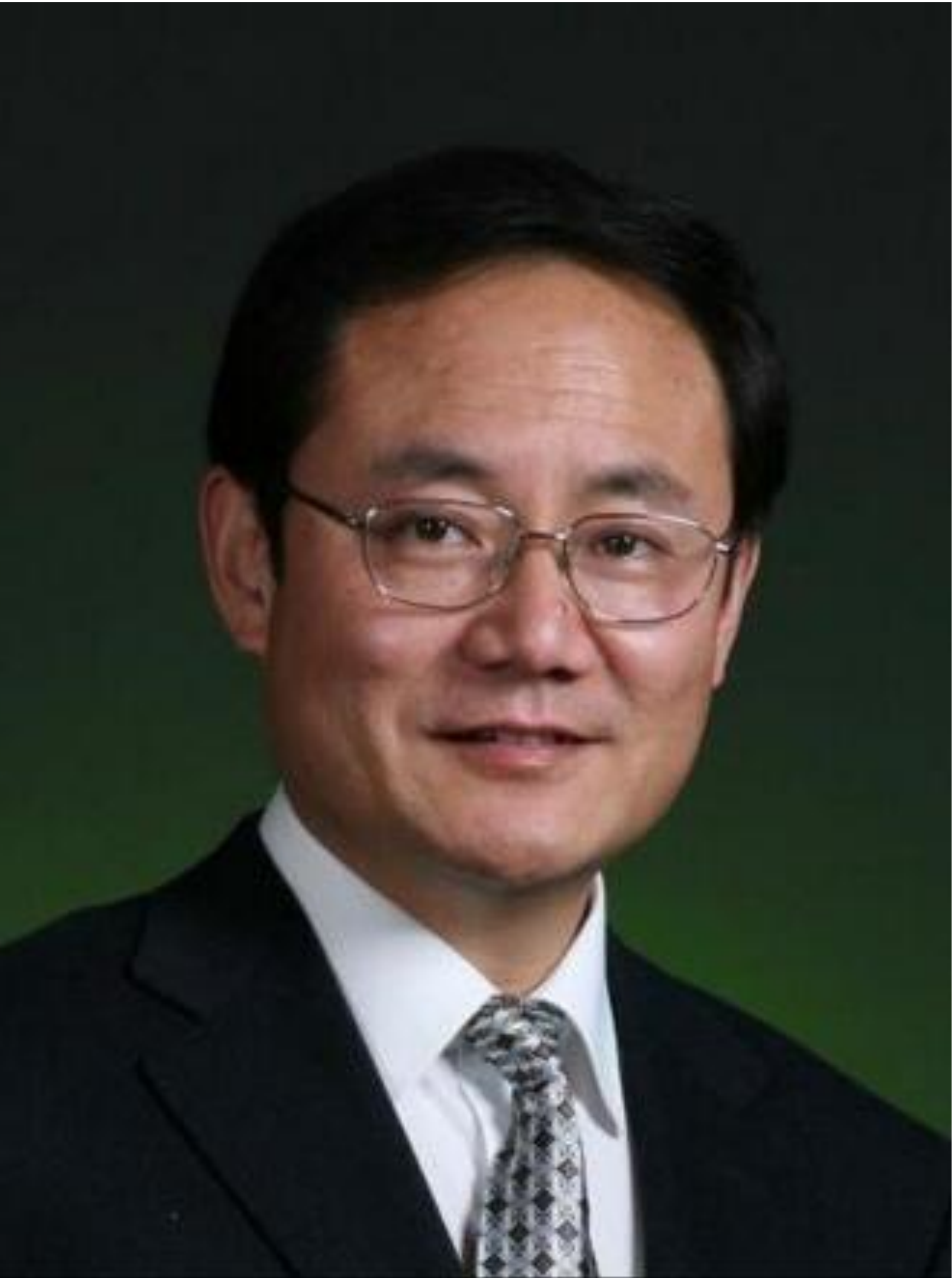}}]{Fangchun Yang}
received his Ph.D. degree in communication and electronic system from the Beijing University of Posts and Telecommunication in 1990. He is currently a professor at the Beijing University of Posts and Telecommunication, China. His research interests include network intelligence and communications software. He is a fellow of the IET.
\end{IEEEbiography}

\end{document}